\documentclass[useAMS,usenatbib]{mn2e}

\usepackage{graphicx}
\usepackage{natbib}
\usepackage{longtable}

\usepackage{txfonts}
\title[Virgo Cluster Dwarf Galaxies]{Probing the Low Surface Brightness Dwarf Galaxy Population of the Virgo Cluster }
\author[Davies et al.]
{J. I. Davies$^{1}$,
L. J. M. Davies$^{2}$ and 
O. C. Keenan$^{1}$  \\
$^{1}$School of Physics and Astronomy, Cardiff University, The Parade, Cardiff, CF24
3AA, UK. \\
$^{2}$ICRAR M468, University of Western Australia, 35 Stirling Highway, Crawley, WA 6009 \\}

\begin{document}

\date{Original January 2011}


\maketitle


\begin{abstract}

We have used public data from the Next Generation Virgo Survey (NGVS) to investigate the dwarf galaxy population of the Virgo cluster beyond what has previously been discovered. We initially mask and smooth the data, and then use the object detection algorithm $Sextractor$ to make our initial dwarf galaxy selection. All candidates are then visually inspected to remove artefacts and duplicates. We derive $Sextractor$ parameters to best select low surface brightness galaxies using central surface brightness values of $22.5 \le \mu^{g}_{0} \le 26.0$ $\mu g$ and exponential scale lengths of $3.0 \le h \le 10.0$ arc sec to identify 443 cluster dwarf galaxies - 303 of which are new detections, with a surface density that decreases with radius from the cluster centre. We also apply our selection algorithm to 'background', non-cluster, fields and find zero detections. In combination,  this leads us to believe that we have isolated a cluster dwarf galaxy population. The range of objects we are able to detect is limited because smaller scale sized galaxies are confused with the background, while larger galaxies are split into numerous smaller objects by the detection algorithm.  Using data from previous surveys combined with our data, we find a faint end slope to the luminosity function of $-1.35 \pm 0.03$, which does not significantly differ to what has previously been found for the Virgo cluster, but is a little steeper than the slope for field galaxies. There is no evidence for a faint end slope steep enough to correspond with galaxy formation models, unless those models invoke either strong feedback processes or use warm dark matter.

\end{abstract}


\section{Introduction} 
A long standing and critical issue with regard to galaxy formation is the observed large numbers of dwarf galaxies produced in numerical simulations (Kauffmann et al. 1993, Moore et al. 1999, Klypin et al. 1999). This problem refuses to go away, as it is closely tied to the fundamentally hierarchical nature of the proposed galaxy formation process (Stadel et al. 2009).
In their most straight forward form, numerical models performed using the standard $\Lambda$CDM cosmology predict large numbers of dwarfs compared to bright galaxies. Typically there should be approximately $10^{4}$ times more galaxies at five magnitudes fainter than a typical $M^{*}_{g} $ galaxy, where $M^{*}_{g} $ is the characteristic absolute magnitude of the Schechter luminosity function ($-20 \ge M^{*}_{g} \le -19$, Montero-Dorta and Prada 2009).  Current state-of-the art observations of galaxies in the Local Group indicate at most just one order of magnitude more dwarfs of this magnitude, when compared to the two bright galaxies M31 and the Milky Way (McConnachie 2012). Here we define a dwarf galaxy to be one that has a total $g$ band absolute magnitude of $M_{g} \ge -16$, although the problem exists for all galaxies fainter than $M^{*}_{g}$.  The problem is typically parameterised by the faint end slope of the luminosity function with observed values being of order $\alpha = -1.1$ (Montero-Dorta and Prada 2009), while the most straight forward model expectations referenced above predict values approaching the divergent value of $\alpha = -2.0$.

The vast discrepancy between observation and model has led to an increasing effort to both detect more dwarfs and to review the assumptions that underlie the model predictions.

Observationally work has concentrated on exploring the dwarf galaxy population of the Local Group (McConnachie 2012). The prime reason for this is that the absolute magnitude and surface brightness of dwarf galaxies appears to be connected such that fainter dwarfs have lower surface brightness. Having very low surface brightness (LSB) makes them extremely difficult to detect against the relatively bright sky of even the darkest of observing sites. For example, using our above definition of a dwarf galaxy we find that all our detections discussed in this work have central surface brightnesses (CSBs) less than that of the darkest night sky we find here on Earth (about 22.5 $\mu g$, where $\mu g$ indicates magnitudes per sq arc sec). We also know from dwarf galaxies already detected in the Local Group, that some have CSBs a hundred times fainter than this sky background value (McConnachie 2012). Due to their extremely LSB these Local Group dwarf galaxies have been detected, not by their ability to cause a faint 'smudge' on a CCD image, but because they show up as over densities in counts of individual stars i.e. they have only been detected because they are near enough that they can be resolved into stars (Bechtol et al., 2015 and references therein). This is a problem that has limited the detectability of these very LSB dwarfs to distances of order a few Mpc - essentially within the Local Group
\footnote{Note that we may be about to see spectacular progress in our ability to see LSB features and LSB dwarf galaxies beyond the Local Group using telescopes like the Dragonfly Telephoto Array (Merritt et al. 2014).}.
 Spectacular recent work, using stellar counts, has shown that the galactic inventory of the Local Group has grown to about one hundred galaxies - a significant increase in numbers, but one that still does not account for the predictions of the original models referenced above. The jury is still out on just how many individual stellar systems exist in the halos of galaxies like the Milky way.

Looking at the problem from the other side, the theorists have been exploring ways in which they might alter their models and their initial predictions. The most commonly accepted and most explored option is the implementation of 'feedback' into the galaxy formation models. Essentially the models are modified to include 'baryon physics' and hence the very reasonable suggestion that stellar winds, primarily due to supernovae explosions, expel gas from dwarf galaxies because of their relatively small gravitational potential compared to the explosive energy. Thus whereas brighter galaxies are able to retain gas for continued star formation, dwarf galaxies lose their gas after the first generation of supernovae - locking in a stellar population made from only a small fraction of the original gas. In addition, as dwarf galaxies are thought to primarily form first in the Universe they are also subject to re-heating of their gas during re-ionisation, thus diminishing the supply of gas for star formation (see for example Kauffmann et al. 1993, Somervile et al. 2001, Benson et al. 2002). Thus both total stellar mass and luminosity are low and the stars are dispersed amongst a disproportionally  large dark matter halo making it a LSB object. Different 'strengths' of the feedback and ionisation effect lead to different derived values of the faint-end slope of the luminosity function within the simulations, showing that it is possible with such a mechanism to make observation ($\alpha \approx -1.2$) match up with theory (see for example Stoehr et al. 2002, Kazantzidis et al. 2004). Hence, the predicted dark matter haloes exist, but we are just not able to detect them. 

\begin{figure*}
\centering
\vspace{-5.0cm}
\includegraphics[scale=0.7]{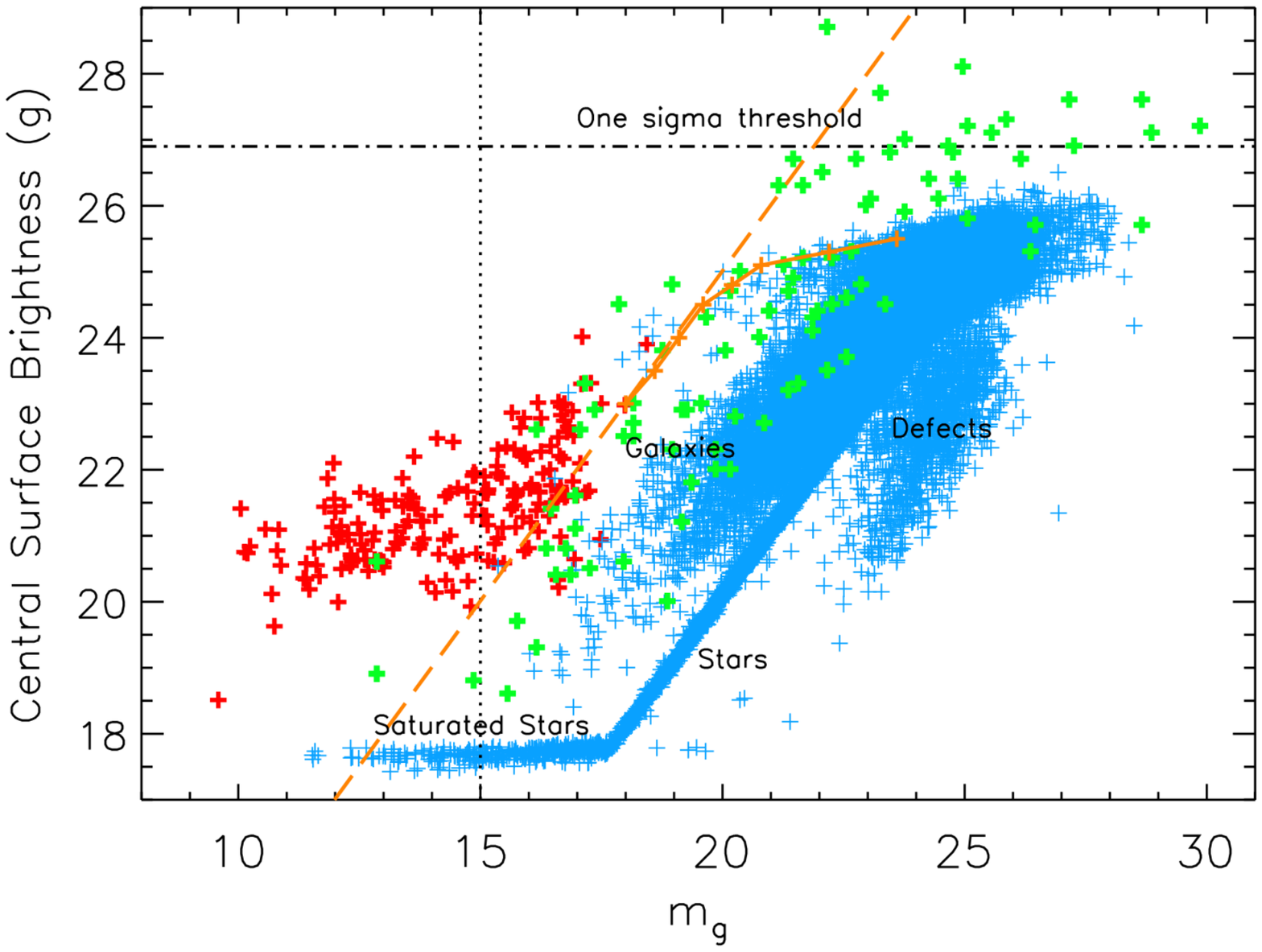}
\vspace{-5.0cm}
\caption{The blue crosses come from running $Sextractor$ on the NGVS+2+1 data frame (see Ferrarese et al. 2012) using the maximum surface brightness parameter as a proxy for the central surface brightness and the isophotal magnitude parameter for the magnitude. There are 134974 objects detected in this un-smoothed frame with a size limit of 30 pixels at a 1$\sigma$ detection threshold of 26.9 $\mu g$. The different types of objects detected are labeled with the vertical dotted line indicating the approximate separation between dwarf and other galaxies at the distance of the Virgo cluster - dwarf galaxies will lie to the right of this line. The red crosses  correspond to VCC galaxies (Binggeli  et al. 1985) that have SDSS data available in the $GOLDMINE$ data base.  We have converted the SDSS data to CFHT magnitudes using $g_{CFHT}=g_{SDSS}-0.15(g_{SDSS}-r_{SDSS})$ and taken the central surface brightness to be 1.7 magnitudes brighter than the average value we can calculate using the given apparent magnitude and size. The green crosses are data for 93 galaxies in the Local Group taken from (McConnachie, 2012). We have adjusted from absolute to apparent magnitudes using a distance modulus of 31.2. We have converted from the given V band magnitudes using $g_{CFHT}=V-0.15(g_{SDSS}-r_{SDSS})+0.28$, but as we can only find values of $(g_{SDSS}-r_{SDSS})$ for 18 of the galaxies we have used the mean value of $<(g_{SDSS}-r_{SDSS})>=0.46$ to do the conversion between magnitudes. The orange dashed line indicates the position of a galaxy with an exponential scale size of 4 arc sec. The solid orange line shows how the values measured by $Sextractor$ for the maximum surface brightness parameter differ from the true central surface brightness value.}
\end{figure*}

However, there are still problems with this scenario, particularly the fact that many dwarf galaxies have been shown to have  star formation histories that are quite complex with both ancient and recent star formation - something that is inconsistent with a early 'strong'  feedback mechanism that essentially switches star formation off (Weisz et al. 2014). The thinness of some galactic discs also remains a problem as the continued existence of numerous dark matter haloes plunging through it increases the velocity dispersion of the stars and hence the scale heights they are able to reach. More elaborate cosmological solutions to the dwarf galaxy problem have invoked 'warm' rather than 'cold' dark matter. This suppresses the formation of the numerous small dark matter haloes so that neither the dwarf galaxy nor the dark halo exist (Lovell et al. 2014). The true nature of dark matter will be tested extensively with the next generation of large volume surveys, such as with the Square Kilometre Array (Power et al. 2015) and the Wide Area Vista Extra Galactic Survey (WAVES,  Driver et al 2015). 

Statistical inference using a sample of one is a dangerous game. Thus it is of prime importance to try and investigate the dwarf galaxy population in regions of the Universe beyond the Local Group. Derivations of the luminosity function down to our prescribed dwarf galaxy regime have been made using both SDSS (Blanton et al. 2003, Montero-Dorta and Prada 2009) and the 2dF redshift survey (Norberg et al. 2002). Both these surveys agree on a faint-end slope to the luminosity function of $\alpha \approx -1.2$, but both, because of their nature, suffer from severe surface brightness selection effects. These affect both the initial selection of objects to those with typically CSBs of less than 23-24 $\mu g$ and probably more importantly, the inability to obtain spectra for objects with CSBs of less than about 23 $\mu g$. As such, these surveys have not explored the possibility of a large population of LSB dwarf galaxies and how it may influence our inference of the value of the luminosity function faint end slope.

Typically, previous optical surveys, such as SDSS and 2dF have been 'shallow and wide' and so were not designed for, and hence not capable of, finding LSB galaxies. On the other hand, deep cosmological surveys have the potential to explore for LSB objects, but as their area coverage has generally been very small the local volume over which they could detect such objects has been limited. Thus all sky surveys have been limited by surface brightness, while cosmological surveys by magnitude (restricted to a small local volume). New deep optical 'All sky' surveys such as those carried out by the Dark Energy Survey and the accumulated survey of the Large Synoptic Survey Telescope offer the opportunity to explore the LSB Universe beyond the Local Group and although we cannot hope to match what can be done using individual star counts, we can hope to explore the LSB Universe further than has been done before.

In this paper we consider ways in which we may disentangle the local (within about 20 Mpc) LSB dwarf galaxy population from the huge numbers of background galaxies that are detected in deep CCD images. This is important as for most LSB dwarfs in deep data, the possibility of obtaining a redshift is both impractical, given the large numbers of background objects the few dwarfs can hide within, and impossible given their surface brightness compared to that of the sky.

Although, we do not currently have access to the deep all sky surveys mentioned above there are deep surveys we can use to hone our tools. One such survey is the CFHT observations of the Virgo cluster - The Next Generation Virgo Survey or NGVS (Farrarese et al. 2012). This survey (described in more detail below) covers an area of about 100 sq deg roughly centred on the giant elliptical galaxy M87 at the centre of the nearby Virgo cluster. If a high density of bright galaxies indicates a high density of dwarf galaxies then this is should be an ideal data set to use. In addition, Virgo has a large number of previously identified LSB systems with which to validate our selection technique.

At 17-23 Mpc the Virgo cluster is the nearest large grouping of galaxies to us. This 6 Mpc range corresponds to the distance of its two constituent sub-clusters, with sub-cluster A at $\sim$17 Mpc and B at $\sim$23 Mpc (Gavazzi et al. 1999). The cluster has played a prominent role in astronomical research since the identification of an excess of nebulous objects in this area of sky was noted by both Messier and Herschel. It was first recognised as a group of extra-galactic objects by Shapley and Ames (1926).  The proximity of the Virgo cluster enables us to study both the general properties of galaxies and the way in which the cluster environment may have affected how galaxies evolve. The cluster contains a wide morphological mix of galaxies that subtend some of the largest angular sizes for extra-galactic objects (M87 subtends 7 arc min, and the spiral M58 6 arc min in diameter) and so this allows us to study not only large numbers of galaxies, but also individual galaxies in detail (Davies et al., 2014).

Recent surveys of the Virgo cluster include X-ray (Bohringer et al. 1994), Ultra-violet (Boselli et al. 2011), optical (VCC, Binggeli et al. 1985, ACS, Cote et al. 2004, VGVS, Mei et al. 2010, SDSS, Abazajian et al. 2009, Kim et al. 2014), near infrared (2MASS, Skrutskie et al. 2006, UKIDSS, Warren et al. 2007), far-infrared (IRAS, Neugebauer et al. 1984, Herschel, Davies et al. 2012) and 21cm, (ALFALFA, Giovanelli et al. 2005, VIVA, Chung et al. 2009, AGES, Taylor 2010). 

Prominent amongst these surveys is the optical survey of Binggeli et al. (1985), which listed 2096 objects over roughly the same area of sky as the NGVS with the resultant list becoming known as the Virgo Cluster Catalogue (VCC). Although the VCC has been shown to include some galaxies that are in the background and many that have no definitive confirmation of membership (see below) it has subsequently served as the primary input for many of the other surveys of the cluster and as a list of its individual galaxy population. The VCC has a stated B band completeness limit of 18.0, which for a distance of 17 Mpc, which is the distance we will adopt here, gives a distance modulus of 31.2 and a minimum dwarf galaxy absolute magnitude limit of $M_{B}=-13.2$. A further important issue is the limited range of surface brightnesses that the VCC photographic data can explore. As we will show below, if we are to detect fainter dwarf galaxies than those in the VCC we will need to extend the surface brightness range of the VCC beyond its sensitivity limit of about 24 B$\mu$. 

The most extensive optical survey of the Virgo cluster since the VCC was created, is that by Kim et al. (2014). They have used SDSS data to select a sample of galaxies, which either have redshifts that confirm their Virgo membership ($v<3000$ km s$^{-1}$) or have a morphology that indicates Virgo membership. Kim et al. observed a larger area than Binggeli et al. (1985) did for the VCC as they were also interested in investigating the extended distribution of galaxies around the cluster - the data is contained in the extended VCC or EVCC. 

Clearly, no other galaxy cluster has had its galaxy population investigated to this level of detail and for this reason we now intend to try and extend the VCC and EVCC further into the dwarf galaxy regime. 

Using the deep NGVS data and a new selection technique we aim to increase the number of known faint low surface brightness galaxies in the Virgo cluster. We will describe why the NGVS data is ideal for this work and how it can be smoothed to improve its surface brightness sensitivity (section 2). We will then show how we can adjust the default parameters of the commonly used image detection program $Sextractor$ so that LSB features are retained - rather than splitting them into numerous small objects - and provide an objective selection method. We then implement a further selection criteria to separate cluster galaxies from the large numbers detected in the background (section 3). Of crucial importance is an analysis of what we might have missed in this survey (section 4). Finally we discuss the numbers of objects detected, the implications of our selection methods and how this influences inferences made about the luminosity function faint end slope (section 5). 

\section{The NGVS data}
The NGVS is comprehensively described in Ferrarese et al. (2012), so here we will just review the important aspects of the data that are relevant to our project. The data has been taken using MegaCam on the Canadian France Hawiian Telescope (CFHT). We use the $g$ band data, which has a similar transmission function to the SDSS $g$ band that has become a standard. Ferrarese et al. give the following relation between the CFHT $g$ band and the SDSS $g_{CFHT}=g_{SDSS}-0.153(g_{SDSS}-r_{SDSS})$, which for a typical galaxy colour of $(g_{SDSS}-r_{SDSS})=0.6$ (James et al., 2008) gives $g_{CFHT}=g_{SDSS}-0.09$ - they are very similar. We will also compare and convert between the VCC B band data for which the SDSS website gives the following relationship $g_{SDSS}=V+0.6(B-V)-0.12$, which again for an approximate galaxy colour of $(B-V)=1.0$ gives $g_{SDSS}=B-0.52$ or $g_{CFHT}=B-0.61$.

\begin{figure}
\vspace{-2.0cm}
\centering
\includegraphics[scale=0.43]{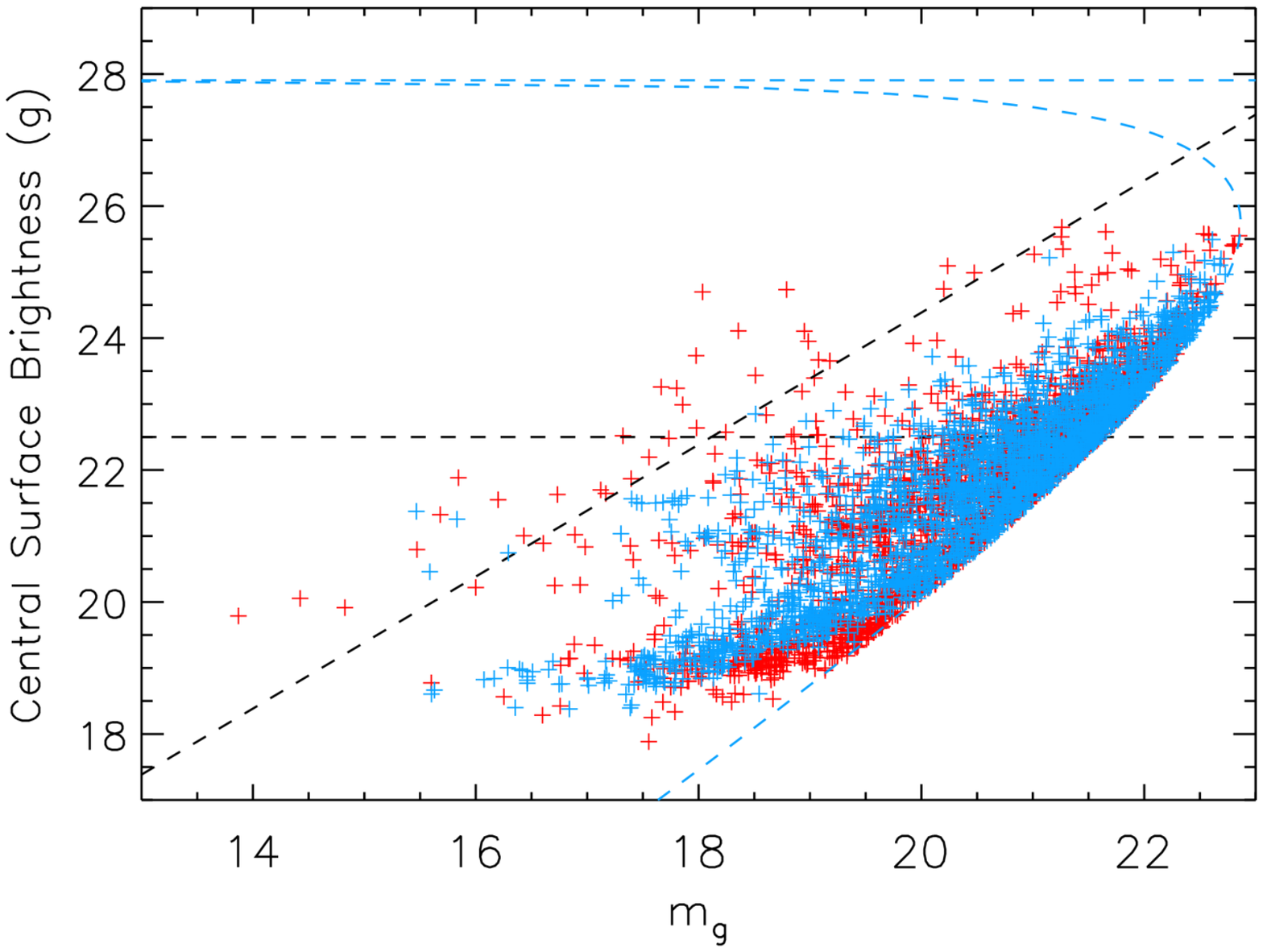}
\vspace{-3.5cm}
\caption{The difference between the properties (magnitude and central surface brightness) of galaxies detected in a cluster field (red, NGVS+0+1) and a background field (blue, BG3).  We have selected only those objects with an isophotal area of 800 pixels (if circular a radius of 3 arc sec) as bounded by the curved blue dashed line. The 1$\sigma$ surface brightness limit of this smoothed frame is 27.9 $\mu{g}$ indicated by the upper blue dashed line. The two black dashed lines indicate the selection criteria of a central surface brightness of greater than 22.5 $\mu{g}$ and an exponential scale size greater than 3 arc sec. Only galaxies on the cluster 'frame' (red) satisfy this criteria i.e. they lie above both black dashed lines.}
\end{figure}

The NGVS consists of 117 pointings covering 104 sq deg with a cluster coverage that is very similar to that of the VCC (see Fig 2, Farrarese et al. 2012). Each frame covers $\approx 1$ sq deg, but adjacent fields overlap by 3 arc min. In addition to the Virgo fields there are four 'background' fields that were designed to lie at $3 \times R_{Virial}=16^{o}$ away from the cluster centre - $R_{Virial}$ is the Virial radius of the cluster. These fields will play an important role in assessing the background contamination in the data we retrieve from the cluster fields. Fig 4 of Farrarese et al. (2012) shows the positions of the cluster fields and Fig 5 the location of the 'background' fields. Each $g$ band data frame is the combination of five dithered 634 second exposures with a minimum acceptable seeing FWHM of $\leq 1$ arc sec and a maximum lunar illumination of 10\%.

All the fields have been processed (data reduction and calibration) at CFHT using the $Elixir$ package as described extensively in Fararese et al. (2012). The result according to Fararese et al. is a signal-to-noise(=10) point source  limit of $m_{g}=25.9$ and an extended source surface brightness limit of $\mu =29.0$ $\mu g$, indicating the high quality and depth of the data compared to what has been available before. We will re-derive surface brightness limits applicable to our methods of detection in the following section. For our use and analysis we have downloaded all 117 $g$ band cluster and 4 background fields from the CFHT archive.

\section{Object detection}
Our intention is to use an objective detection method to select cluster dwarf galaxies on the NGVS data frames. By objective selection we mean that the criteria for selection are well defined so that we can realistically assess the significance of what we have found and, probably more importantly, what we have not be able to find. As described below our selection criteria is based on both a size and surface brightness limit. Although Fararese et al. (2012) state a surface brightness limit of $\mu=29.0$ $\mu g$  it is not clear over what area this applies. Through our own analysis we measure a 1$\sigma$ pixel-to-pixel noise of about 26.9 $\mu g$ in each frame. It is this number that is important as we will be using the object detection algorithm $Sextractor$ to detected sources as groups of connected pixels that all lie above a threshold which is some multiple of this pixel-to-pixel noise. 

\subsection{Using $Sextractor$}
Before implementing the above process we have reduced the original pixel-to-pixel noise by smoothing the data using a Haar filter (Davies et al., 2005). The Haar filter is a form of wavelet that enables us to smooth the 'sky' between bright objects without smoothing the bright objects themselves. When applied to the data the transform produces adjacent pixel difference values that, if assumed to be due to random noise can be set to zero, while those that are due to real gradients in the data are retained. Upon image reconstruction the 'random' noise is reduced. The result of this mild Haar smoothing of the data is that the originally measured 1$\sigma = 26.9$ $\mu g$ is increased by about one magnitude to typically 27.9 $\mu g$. 

Experimentation using $Sextractor$ highlighted three issues that are specifically relevant to using a connected pixel algorithm to detect LSB objects (see also Sabatini et al. 2003):
\begin{enumerate}
\item There are large numbers of false detections caused by the faint LSB halos of bright stars and by satellite trails across the data. To overcome this problem we have automatically masked bright stars. This typically reduces the detection area to about 85\% of what it was originally. We have also imposed a limit on the measured ellipticity of objects detected at $e \le 0.4$, where $e=1-b/a$, hence we are selecting 'round' objects - 'detected' objects in satellite trails, a significant source of false detection, are long and thin and so are also excluded. Our intention is to try and detect the most significant Virgo dwarf galaxy population, which we expect to be made up of spheroidal ($e \approx 0.0$) dE galaxies. For disc galaxies they would satisfy this condition on ellipticity as long as their inclinations to the line of sight are less that 66$^{o}$, which for randomly orientated discs will happen about $\approx$60\% of the time. As such, we will miss about 40\% of LSB dwarf disc galaxies, which have higher inclinations. \\
\item As $Sextractor$ was designed to identify faint, often point like, sources it has software to separate fluctuations in brightness into individual objects. The problem when detecting LSB objects is that random noise can lead to areas of the the object appearing brighter than others and so it is split up by $Sextractor$ into a number of smaller components. To reduce, but not cure this problem (see below), when carrying out our detection, we have switched off the deblending option. \\
\item As we will describe below our selection process requires both a measurement of surface brightness and exponential scale length. $Sextractor$ produces a value for the maximum surface brightness in the image. The problem with this maximum value for a LSB object is that it is governed by the noise in the image rather than anything to do with the detected galaxy. For example, for Gaussian noise the probability that a pixel has a value greater than $\approx3\sigma$ is about 0.001. For an image consisting of 800 pixels (our initial minimum size limit, see below) we might expect one pixel of this value just by chance. 
In Fig. 1 we show why this is a problem when using this maximum surface brightness value to measure the true surface brightness of LSB galaxies. The $Sextractor$ maximum surface brightness pixel (see orange line and figure caption) corresponds very well with the simulated central surface brightness (CSB) value until about 24.5 $\mu g$ at which point the measured value tends to a constant value of $\approx$25.7 $\mu g$ i.e. determined by the noise rather than the actual CSB of the galaxy. An alternative to this surface brightness value is to form a hybrid surface brightness from the magnitude and size of each galaxy, for example $<\mu>=m_{ISO}+2.5\log{\pi R_{ISO}^{2}}$, where $m_{ISO}$ and $R_{ISO}$ are the isophotal magnitude and radius respectively - these values are also calculated by $Sextractor$. The problem here is that this measure of surface brightness is very insensitive to galaxies of intrinsically different surface brightnesses, because as the central surface brightness is reduced, the area over which $M_{ISO}$ is measured gets smaller with the reduction in $R_{ISO}$. Simulating a 3 arc sec galaxy (exponential profile) shows that a change in CSB of 3.5 mag arc sec (24 to 27.5) results in a 1.6 mag arc sec change in $<\mu>$. Thus galaxies of a wide range of intrinsic surface brightness get 'mixed up' into a narrow range of measured surface brightness. To overcome this problem we have used the isophotal values that $Sextractor$ produces to retrieve the CSB based on the assumption that the galaxies have exponential profiles - this is the CSB consistent with $m_{ISO}$ and $R_{ISO}$. This greatly improved our detection method with very few intrinsically LSB galaxies being detected on the 'background' frames, while many LSB galaxies were revealed on the cluster frames (see below).
\end{enumerate}

\subsection{Expected properties of Virgo cluster dwarf galaxies}
Our next concern was the photometric properties of the galaxies one might expect to detect in the Virgo cluster. We have investigated this in two ways. Firstly, we can extrapolate what is already known about the Virgo cluster dwarf galaxy population and secondly we use the properties of galaxies in the Local Group to predict what we might find. On Fig. 1 we have plotted object magnitude against CSB ($g$ band) for objects detected on a single example NGVS frame (light blue) compared to known Virgo cluster galaxies (red) and known Local Group dwarfs (green). To create Fig. 1 we have run $Sextractor$ on a NGVS frame with a small pixel area selection limit (30 pixels - one pixel is 0.19 arc sec), so that we detect a variety of objects (as indicated on Fig. 1) - for these objects we plot the isophotal magnitude against the maximum surface brightness value as given by $Sextractor$. This is a data frame before Haar filter smoothing and so illustrates the limiting faint surface brightness cut-off discussed above i.e. the limit in measured maximum surface brightness. 

For comparison with the stars, galaxies and defects (bad pixels, single pixel 'events' etc) detected on the NGVS frames we have also plotted on Fig. 1 data  for known VCC galaxies. We have taken this data from the $Goldmine$ database (Gavazzi et al. 2003), but due to the limited availability of data for comparison we have had to make some approximations. The $Goldmine$ data comes from SDSS and so we use the conversion $g_{CFHT}=g_{SDSS}-0.153(g_{SDSS}-r_{SDSS})$ (Farrarese et al. 2012) so that we can compare magnitudes. Also there is no measure of surface brightness in the $Goldmine$ database, only a size measured at 'the last visible isophote'. We have used this to calculate a mean surface brightness $< \mu > =m_{g}-5.0\log{a}$, where $m_{g}$ is the apparent magnitude and $a$ the area at the 'last visible isophote'. For an exponential profile the mean surface brightness over the half-light radius is 1.7 mag fainter than the CSB so we have assumed that $<\mu>$ is this mean value and made the 1.7 mag sq arc sec adjustment to get a value for the CSB. Although this is a crude means of calculating the central surface brightness, we feel that it is sufficiently good to be able to compare the surface brightness of known Virgo cluster galaxies with what we might expect for lower luminosity dwarf galaxies. The data is plotted as red stars on Fig. 1. Clearly if the relation between CSB and magnitude persists to fainter magnitudes then the dwarf LSB galaxies will quickly merge into the large numbers of other faint background galaxies on the frame. 

For comparison with the now numerous dwarf galaxies detected in the Local Group we have used the compilation of data given in  McConnachie (2012), which is primarily in the V band.  As explained in the caption to Fig. 1 we have converted between the V band and $g_{CFHT}$, so that we can compare with the VCC and NGVS data directly. We have also adjusted from absolute to apparent magnitudes using a distance modulus of 31.2. If a relationship exists between magnitude and surface brighteness for Local Group galaxies then clearly it is a different relationship than that for the VCC galaxies (the green points do not sit in the same region of parameter space as the red points). Many of the Local Group galaxies are of too LSB to be detected on the NGVS frames. However, some could be detected, but then the issue would be how to distinguish them from the large numbers of background galaxies. So, our problem is how to separate out those in the Virgo cluster from those that are intrinsically brighter galaxies, but further away (with similar apparent magnitudes and angular sizes).

\begin{figure*}
\vspace{-2.0cm}
\centering
\includegraphics[scale=0.42]{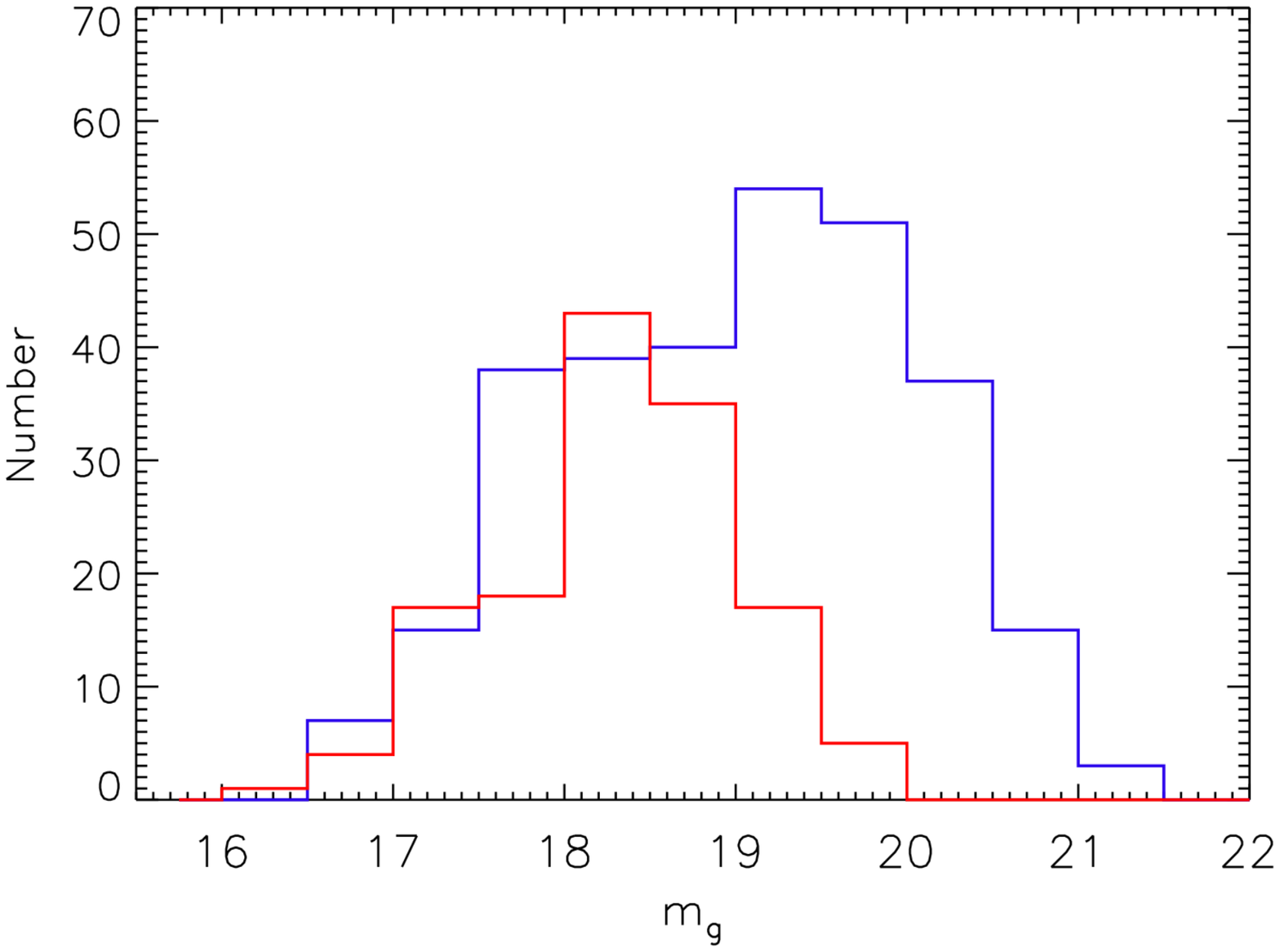}
\includegraphics[scale=0.42]{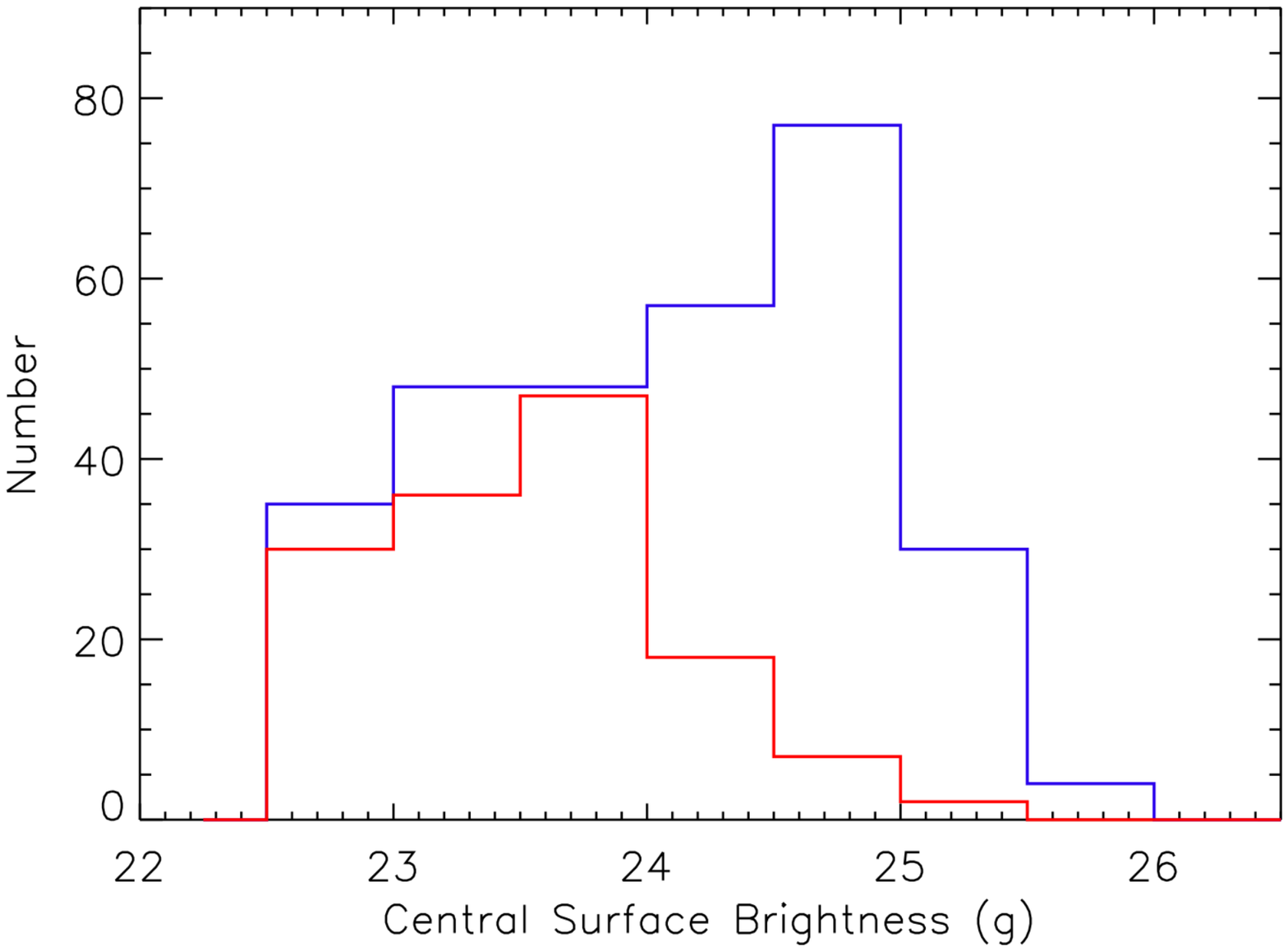}
\vspace{-3.0cm}
\caption{Left - distribution of apparent magnitudes for the 443 galaxies detected. Right - distribution of CSB for the 443 galaxies detected. Red is for galaxies that are also in the VCC and blue for new detections. }
\end{figure*}

\subsection{Deriving the selection criteria}
To try and separate cluster dwarf galaxies from background objects we consider their positions in the magnitude-CSB diagram. Our objective is to devise selection criteria that minimise the numbers of objects detected on the four 'background' frames while maximising the numbers of LSB objects found on the cluster fields. After attempting various different methods we were actually led to exactly the same conclusions as Sabatini et al. (2003). Their selection criteria was an exponential scale size greater than 3 arc sec and a CSB fainter than 23.0 B$\mu$. Using the same scale size selection and an approximately equivalent CSB threshold of 22.5 $\mu g$ 
\footnote{A linear fit to the B against $g$ magnitude for VCC galaxies detected by us gives $g_{CFHT}=B_{VVC}-0.23$ thus 23.0 B$\mu$ is approximately 22.8 $\mu g$ for these galaxies, close to the value of 22.5 $\mu g$ we have used.} we detect no objects on the 'background' fields, but an average of 4 per cluster frame (Fig 2). There is also a clear concentration of detected objects towards the cluster centre (see below). Fig 2. illustrates this selection by showing the distribution of apparent magnitude against CSB for a field close to the cluster centre compared to one of the background fields. 

In summary:
\begin{enumerate}
\item we have selected all objects with an area less than or equal to 800 pixels (if circular a radius of 3 arc sec) and a measured ellipticity of $e \le 0.4$.  
\item we use the measured isophotal magnitudes and areas to calculate a CSB ($\mu_{0}$), scale size ($h$) and total apparent magnitude ($m_{g}$) assuming an exponential surface brightness profile. 
\item we finally select only those galaxies with $\mu_{0} \ge 22.5$ $\mu g$ and $h\ge3.0$ arc sec.
\end{enumerate}

The above selection criteria resulted in a sample of 443 objects in the NGVS data. This is after visual inspection of every object to ensure it was not a defect and the removal of duplicates in the overlap regions of the frames. There were no detections in any of the 4 'background' frames. Comparing our detections with galaxies already listed in the VCC we found 140 duplicates, thus there are 303 new LSB dwarf galaxies in our sample. In Fig. 3 we show distributions of magnitude and CSB for the galaxies in our sample distinguishing them as either known VCC galaxies or new detections. We have obviously been somewhat conservative with our selection in this very much deeper data, because we have only extended both the apparent magnitude and CSB distributions by about one magnitude from those of the VCC, but we are convinced that this is a sample of Virgo galaxies - we will discuss the limitations of our method below. The major problem is defining a region of the CSM - magnitude plane that contains cluster dwarfs but is also free of background galaxies. To do this we have reduced what was originally about 5 million detections to about 400 in the final list.  On the plus side we are detecting objects that at the distance of the Virgo cluster have absolute magnitudes as low as $M_{g}=-10.0$ and CSB of 25.0 $\mu g$.

\subsection{Inferring cluster membership}
As an indication of whether we are detecting cluster rather than background galaxies we can consider the relationship between the surface number density of our detections and their positions in the cluster. In Fig. 4 we have plotted our detections (blue) together with the previously known VCC galaxies (red). Previous observations indicate that the Virgo cluster does not appear to be a single dynamic entity, but rather is made up of a number of different groupings (see Davies et al. 2014 and references therein). Most prominent of these two groupings are the two sub-clusters A and B centred on the two elliptical galaxies M87 and M49 respectively, the positions of these two galaxies are marked on Fig. 4 (left). It is quite clear from Fig. 4 that our new detections (blue) cluster quite strongly around M87, but the area around M49 is rather devoid of them - just why the two sub-clusters are different is not totally clear but we note that  Binggeli et al. (1987) describe sub-cluster A as rich in early type galaxies, while they describe sub-cluster B as being rich in late type galaxies. They also say that sub-cluster B is falling into sub-cluster A from behind. Gavazzi et al. (1999) place sub-cluster B at 23 Mpc and sub-cluster A at 17 Mpc. Sub-cluster B galaxies would have to be intrinsically 0.7 magnitudes brighter than those in sub-cluster A to find their way into our sample. So, the lack of LSB dwarfs in sub-cluster B may either be due to its greater distance or the different morphology of the brighter galaxy population. The latter explanation is more interesting in that it implies an intimate connection between the dwarfs and their larger hosts. This relationship will be further explored in a subsequent paper in which we will consider in more detail the photometric properties and particularly the colours of the dwarf galaxy population.

\begin{figure*}
\vspace{-2.0cm}
\centering
\includegraphics[scale=0.42]{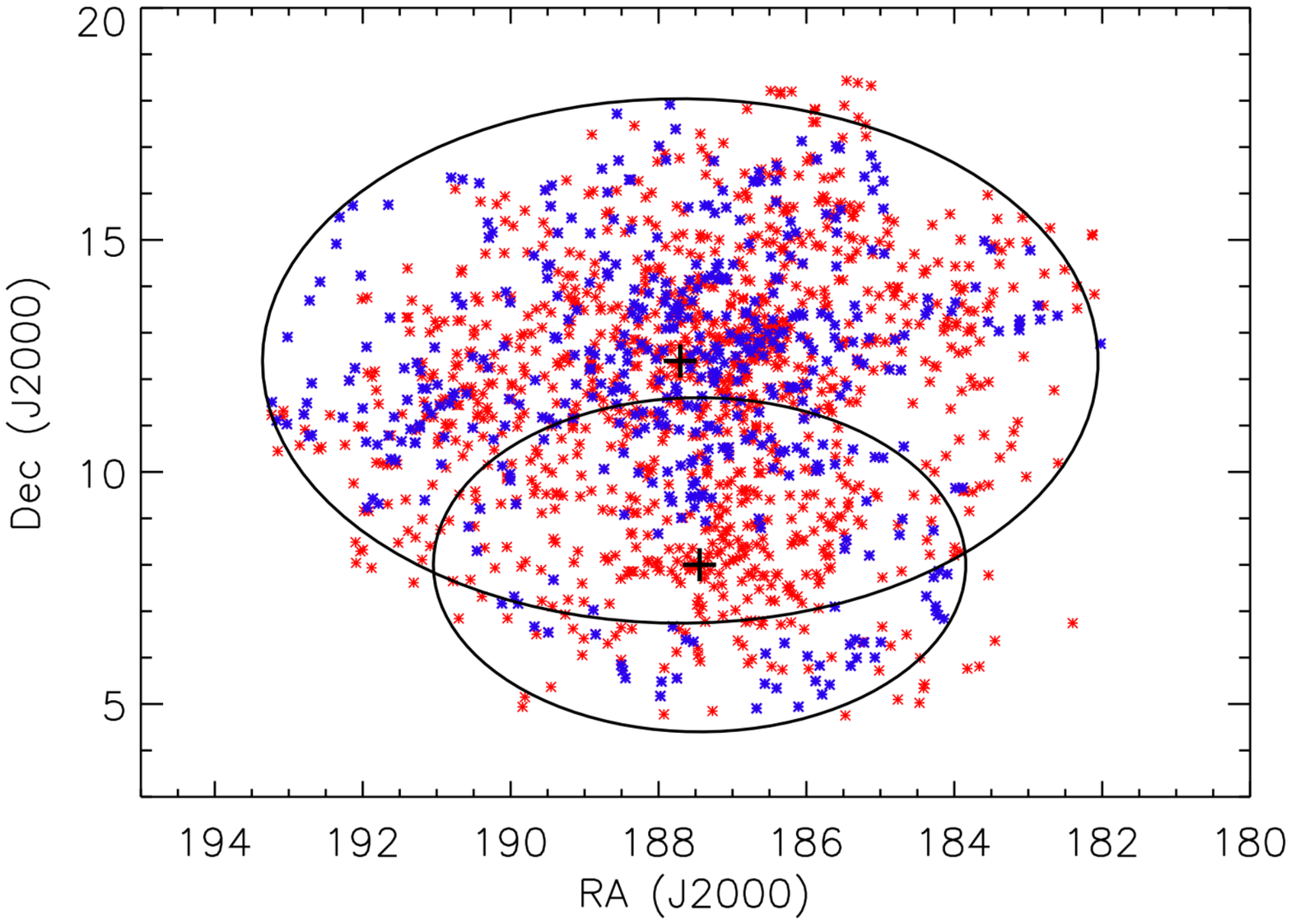}
\includegraphics[scale=0.42]{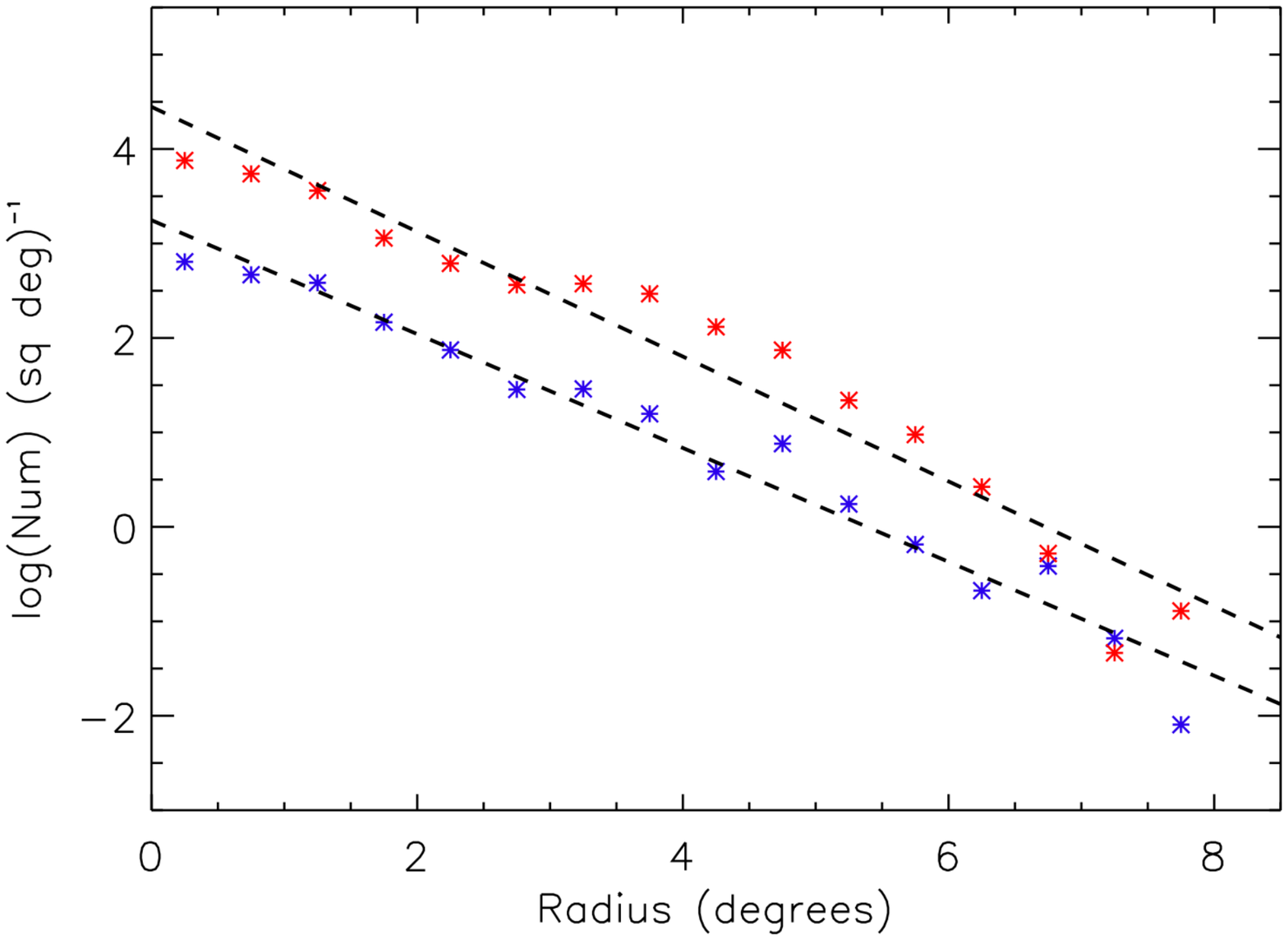}
\vspace{-3.5cm}
\caption{Left - the positions of our detections (blue) compared to all VCC galaxies (red). The black ellipses define the extent of the NGVS data, which is approximately the Virial radius of the two sub-clusters (A and B) centred on M87 (top) and M49 (bottom). Right how the number density of galaxies falls with distance from M87 for our detections (blue) and all VCC galaxies (red).}
\end{figure*}

To quantify the distribution of our new galaxies within the cluster, and to provide further evidence for their cluster location, we have measured their surface number density as a function of cluster centre distance - assumed to be the position of M87 - we have simply counted galaxies in 0.5$^{o}$ circular annuli at increasing radii. The result is shown in Fig 4 (right). Although our new detections clearly avoid some areas of the cluster (Fig. 4, left) on average there is a smooth decline in density with radius consistent with an exponential decline in numbers. We have also compared the observed number density with that predicted for a constant density and an isothermal sphere, but neither of these fitted the observations. A fit to the data gives a central number density of $\approx 26$ galaxies per sq deg and an exponential scale length of 1.7 deg or 0.5 Mpc for a Virgo distance of 17 Mpc. These numbers are very close to those measured by Sabatini et al. (2003) who considered a comparatively small data set consisting of a 1.5$^{o}$ wide strip extending from the cluster centre out to 7$^{o}$ (east to west). They measured a central LSB dwarf galaxy density of about 30 per sq deg and an exponential scale length of 2.2$^{o}$ using similar detection criteria. We have also plotted for comparison on Fig 4. (right) the number density of VCC galaxies. Our new detections follow the distribution of previously detected VCC galaxies remarkably well (exponential scale length of 1.5$^{o}$). Given the clear concentration of these galaxies towards the cluster centre we are very confident that our sample galaxies are members of the Virgo cluster.

\begin{figure*}
\vspace{-2.0cm}
\centering
\includegraphics[scale=0.4]{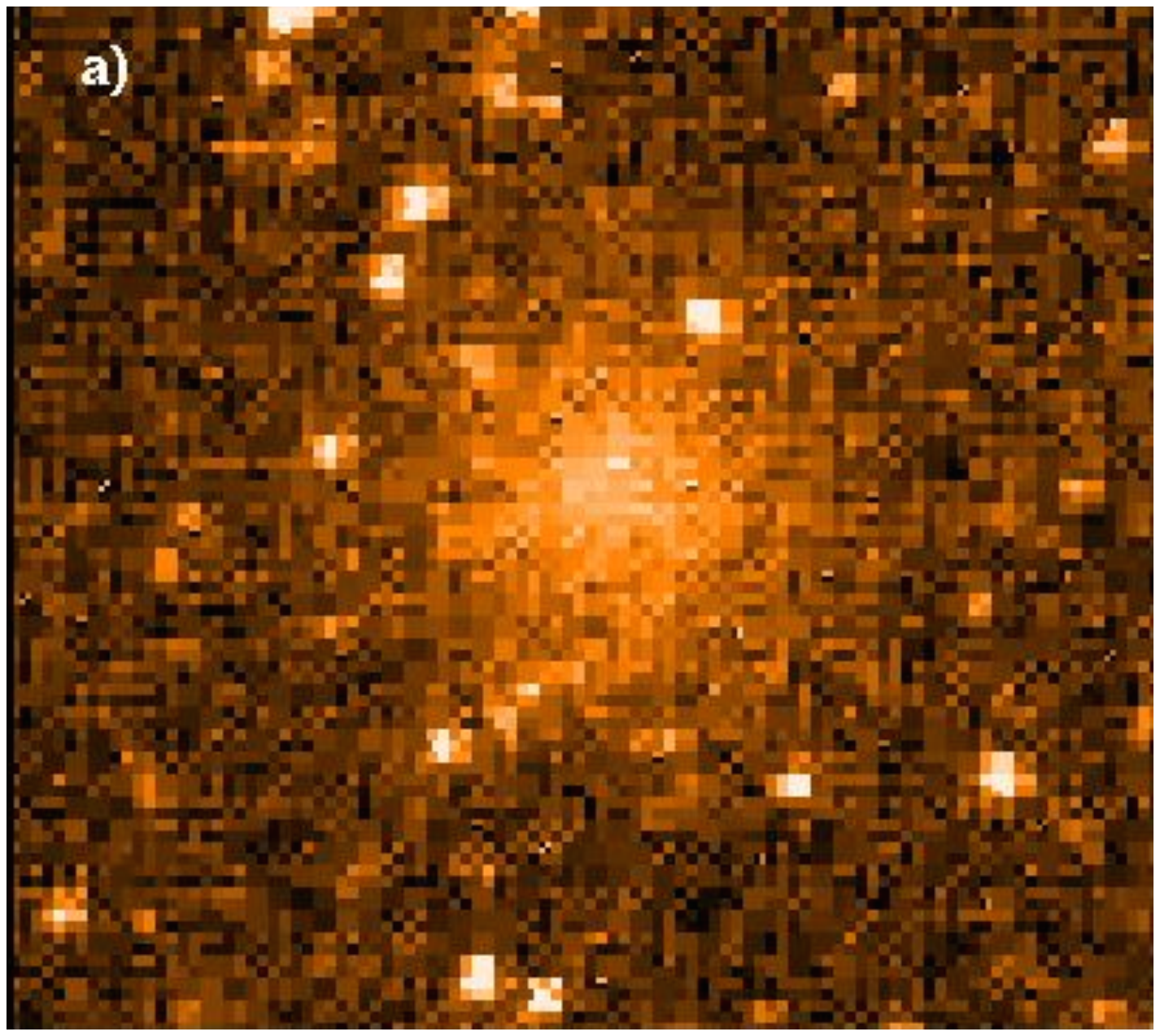}
\includegraphics[scale=0.4]{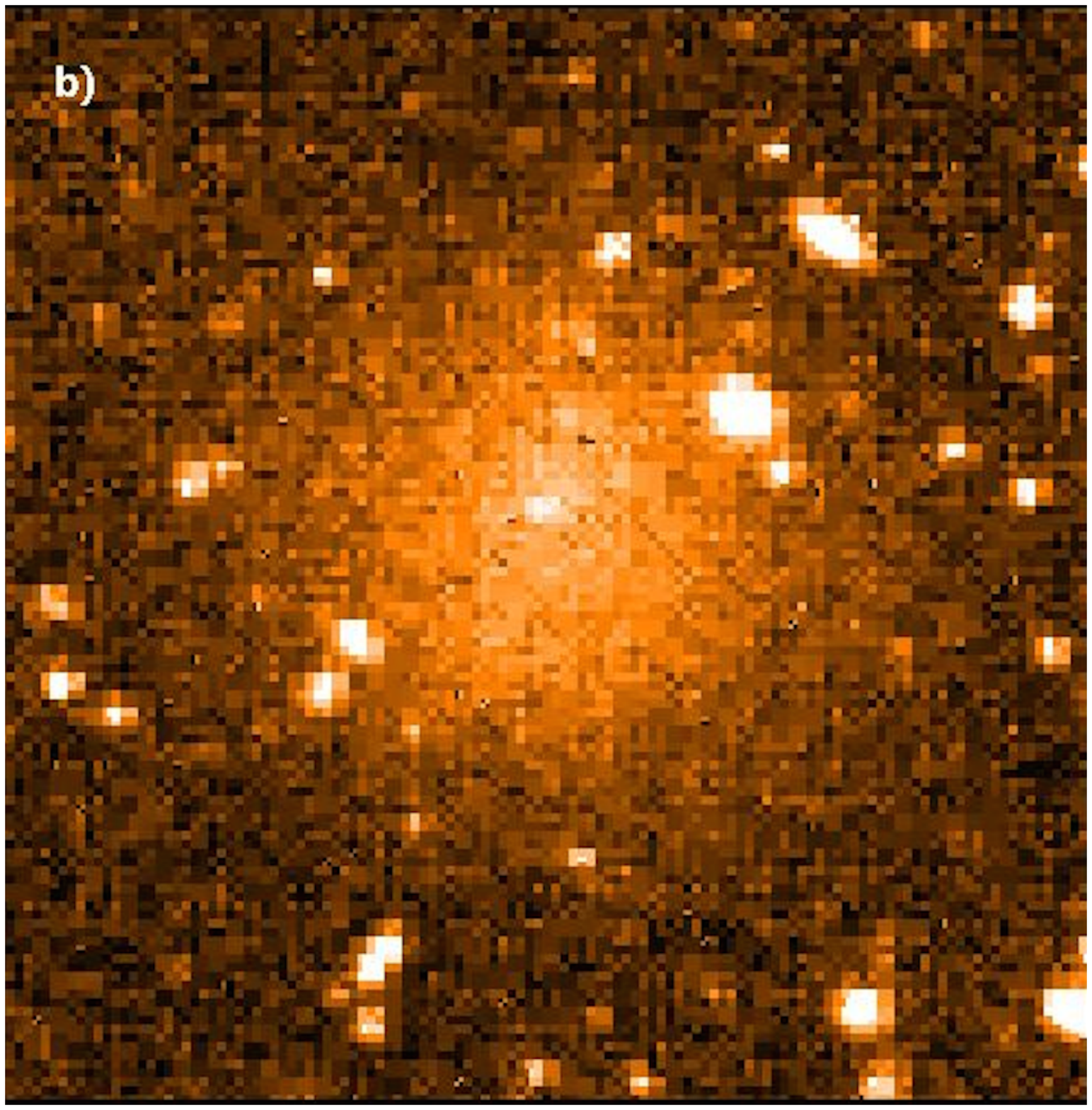}
\vspace{-2.0cm}
\caption{Examples of two galaxies detected by us. a) $m_{g}=20.5$, $\mu_{0}=25.0$ $\mu g$, $h=3.2$ arc sec. b) $m_{g}=19.2$, $\mu_{0}=24.4$ $\mu g$, $h=4.3$ arc sec. The cut-out images are different sizes to illustrate the different sizes of the two galaxies - image a) has a size of $\sim38$ and b) $\sim46$ arc sec, which is 3.1 and 3.8 kpc for a distance modulus of 31.2 (17 Mpc).}
\end{figure*}

\section{Quantifying what we might have missed}
Before moving on to look at how our new detections affect the galaxy luminosity function, we want to consider what objects we may have missed using our selection criteria. Firstly, we will compare our results to what has been found in other surveys, and secondly we will consider the properties of objects that might exist in Virgo, but we could not detect. We will start with two surveys that were aimed specifically at finding LSB galaxies.

\begin{figure}
\vspace{-2.0cm}
\centering
\includegraphics[scale=0.43]{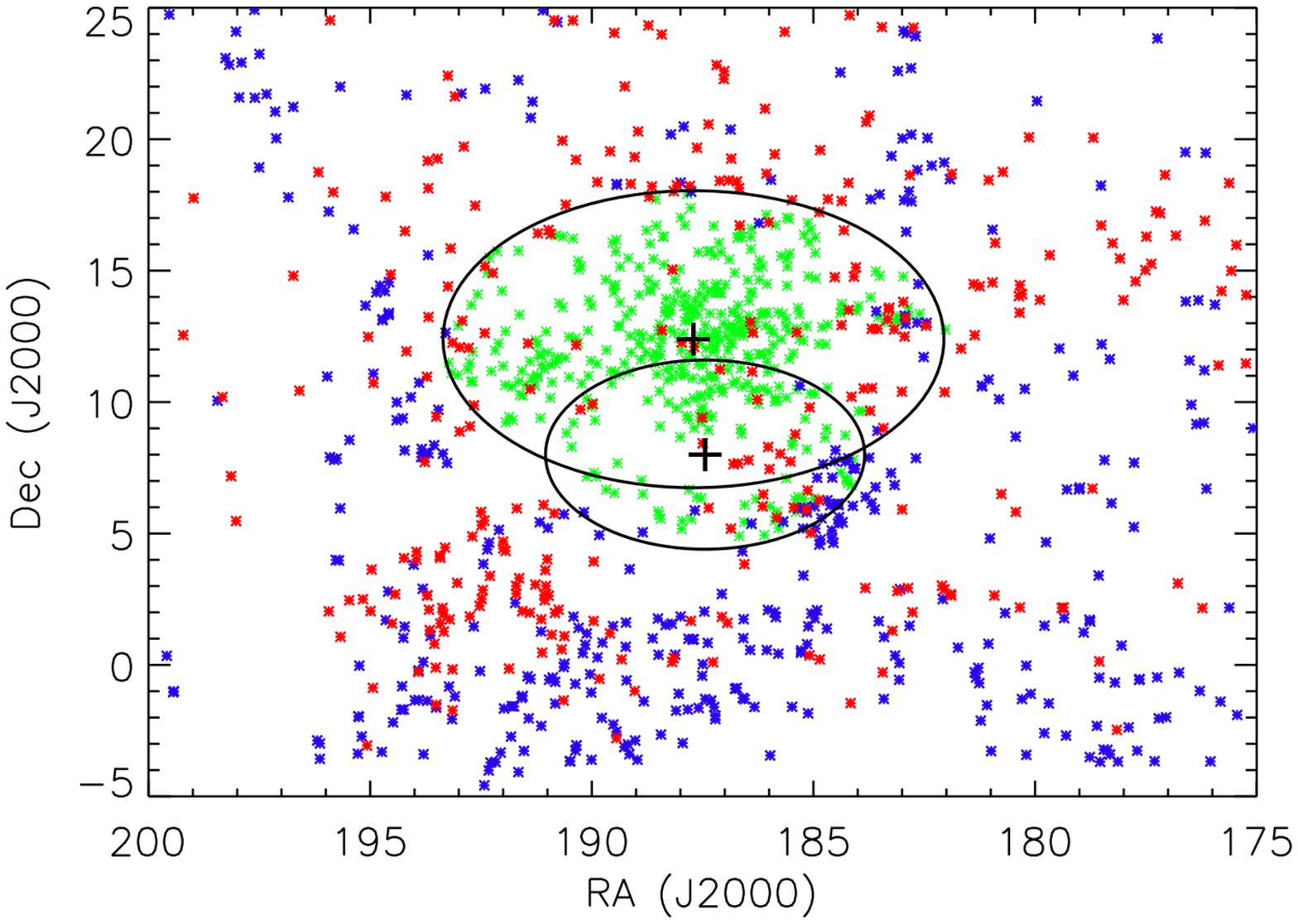}
\vspace{-3.5cm}
\caption{The positions of our detections (green) compared to all 676 newly detected EVCC galaxies (red - certain members, blue possible members). The black ellipses define the extent of the NGVS data, which is approximate the Virial radius of the two sub-clusters (A and B) centred on M87 (top) and M49 (bottom).}
\end{figure}

\begin{figure*}
\vspace{-2.0cm}
\centering
\includegraphics[scale=0.7]{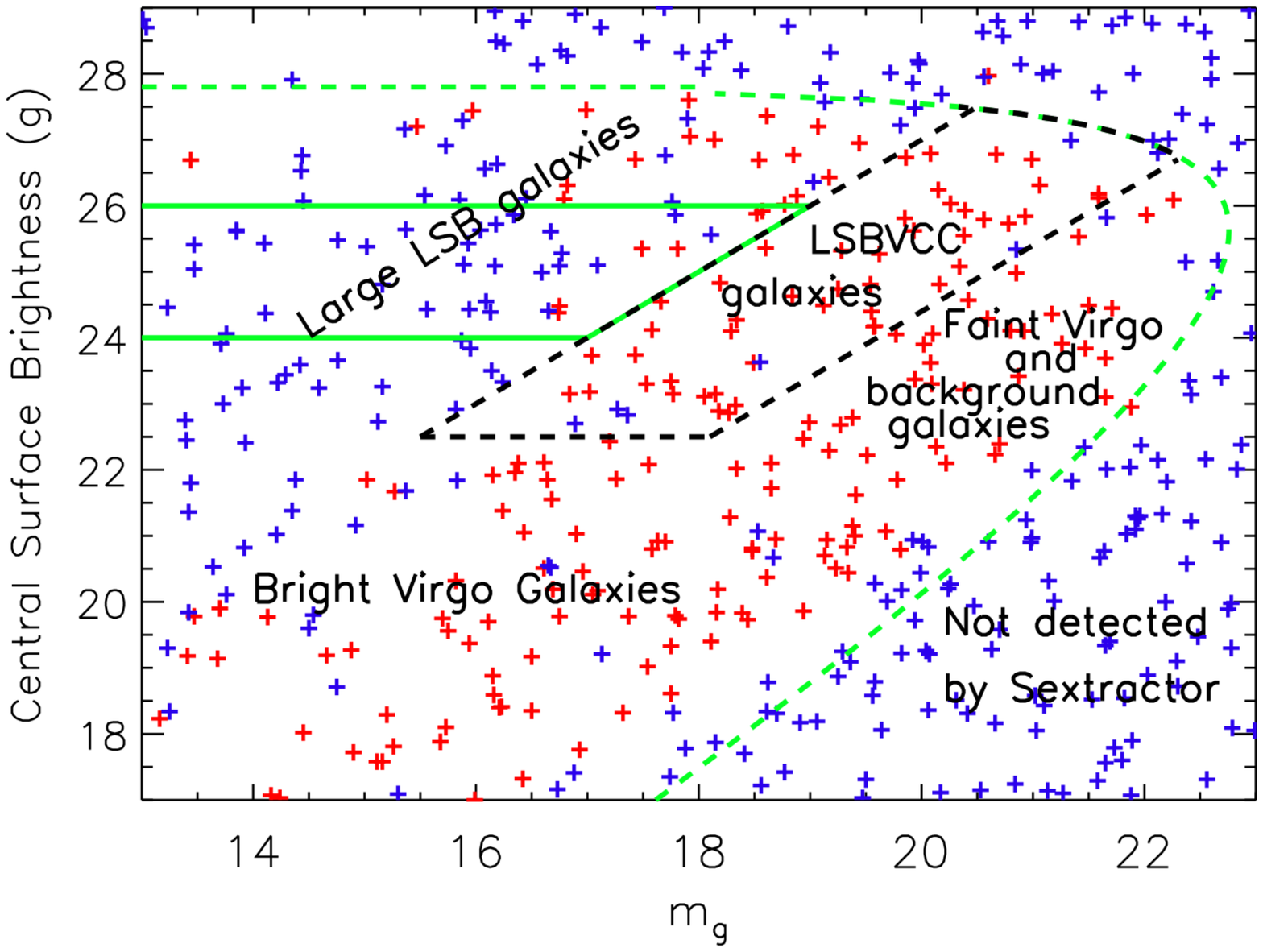}
\vspace{-5.5cm}
\caption{This is the same plot as that shown in Fig. 2 except this is now for simulated data. We have randomly sampled the CSB-magnitude plane and used these parameters to add galaxies to a noise frame. Simulated galaxies that were not detected by $Sextractor$ are indicated in blue, those that were detected are marked in red. The dashed green line indicates the $Sextractor$ initial selection criterion of an area of 800 pixels above 27.8 $\mu g$. The black dashed lines enclose the area of the CSB-magnitude plane we have explored in this paper. The green box (solid lines) indicates the region of the CSB-magnitude plane where van Dokkum et al. (2015) recently found galaxies in the Coma cluster.}
\end{figure*}

The first work to try and extend the VCC to lower surface brightness levels was carried out by Impey et al. (1988). They used Malin's technique of photographic amplification to try and reach lower surface brightness levels - down to a claimed level of 27 B$\mu$ about 1.5 magnitudes brighter than our 1$\sigma$ isophote. The same process was used to identify the giant LSB galaxy Malin 1 (Bothun et al. 1987), which is one of the 137 galaxies they detect over just 7.7 sq deg of sky. Just 27 of these galaxies were new detections, the other 110 being galaxies already listed in the VCC. With the exception of one object that we cannot see at all on the NGVS data, using $Sextractor$ with the parameters described above, we detect all the other 26 objects. However, only 3 make their way into our final catalogue. There are three reasons for this: a number are actually of quite high surface brightness (3 have redshifts that place them behind Virgo), some have values of ellipticity $e > 0.4$, two are actually VCC galaxies (VCC1323 and VCC1235). The 3 galaxies in common have CSB of 25.4, 23.8 and 24.0 $\mu g$, which is certainly pushing the limits of what could have been found by Binggeli et al. (1985) for inclusion in the VCC. We conclude that the Impey et al. (1988) paper is not a source of additional LSB objects that we have missed, other than those that do not satisfy our selection criteria.

Sabatini et al. (2003) considered B band data consisting of a 7 deg long strip extending west to east from the cluster centre (M87) with a total area of about 14 sq deg. They used a convolution method with matched filters (exponential filters with varying scale sizes) to both detect and measure galaxies. The objective was to detect LSB galaxies with CSBs down to 26.5 B$\mu$. The result was the detection of 237 LSB galaxies using, as described above, very similar selection criteria to ours - $\mu_{0} > 23.0$ B$\mu$ and $h>3.0$ arc sec. We do not have the final catalogue of galaxies, but published in Sabatini et al. (2005) are the positions of 15 of the galaxies. All of these 15 galaxies are detected by us with the exception of four of them that lie beyond the extent of our data. Eight of the detected objects are quite bright galaxies and easily detected by $Sextractor$, four have redshifts from 21cm observations with two confirmed as cluster members and two as background. Three objects do not make our sample as their ellipticity is $e > 0.4$. Of particular interest is their object 144 which appears on our frames as an extended very diffuse LSB object. Sabatini et al. measure object 144 to have $m_{B}$=19.8, $\mu^{B}_{0}=26.6$ B$\mu$ and $h = 9$ arc sec i.e. a large scale size combined with a very low CSB. It is detected by us and is included in our sample, but we do not measure such extreme values for it - $m_{g}=19.5$, $\mu^{g}_{0}=24.8$ $\mu g$ and $h = 4.6$ arc sec. We conclude that we are able to detect all the  objects in the  Sabatini et al. (2003) sample using $Sextractor$, but that not all of these objects will be included in our sample either because our data does not extend as far from the cluster centre, because some fail our surface brightness test and because some have $e >  0.4$. 

As described in the introduction the most extensive optical survey of the Virgo cluster since the VCC is that described in Kim et al. (2014). The area observed by Kim et al. is shown in Fig. 6, where new galaxies detected by them are distinguished as being certain (red) and possible (blue) members. The striking feature is their lack of new detections within the Virial radius of the two sub-clusters. Just 110 of the 676 newly detected galaxies reside within the Virial radii (see also the list of SDSS galaxies in this area given in Davies et al. 2014). The galaxies detected by us in this paper are indicated on Fig. 6 in green clearly indicating that the EVCC misses a large number of galaxies within the two sub-clusters Virial radii - just 39 galaxies are in common between our survey and those in the EVCC. Kim et al. do not discuss the selection effects that obviously affect their ability to detect the cluster LSB population. The SDSS survey is rather shallow compared to the NGVS and in addition the requirement of obtaining a redshift means that only higher surface brightness galaxies will be included in the EVCC. It is clear that the VCC contains most of the galaxies in the EVCC over the areas both surveyed. Our survey demonstrates that surface brightness is the real issue and that the VCC can only be extended to fainter magnitudes by deeper surveys that are able to detect much lower surface brightness levels.

Before moving on it is also worth considering some recent work not on Virgo, but the Coma cluster. van Dokkum et al. (2015) used the Dragonfly Telephoto Array to image a $3^{o} \times 3^{o}$ sq deg region centred on Coma. This instrument has the ability to reach impressively low surface brightness levels. The survey identified 47 galaxies with CSB values of 24 - 26 $\mu g$ within the range of surface brightness values found by us in Virgo. However, they also measure large effective radii for these galaxies of between $R_{e}=$ 3 and 10 arc sec. For an exponential profile $R_{e}=1.7h$ and so with a Virgo to Coma distance ratio of 5.8 (using the angular size distance to Coma of 98 Mpc quote by van Dokkum et al.) we would expect galaxies like this to have scale sizes of about 10 - 30 arc sec at the distance of Virgo. The largest scale size we have in our sample is 9 arc sec - so we do not find large LSB galaxies in Virgo like those found in Coma. 

We can investigate whether it is possible to identify such large LSB systems in Virgo using our methods. At the same time we can investigate how complete we are in our ability to detect objects in the region of the CSB-magnitude plane defined by our selection criteria.  We have carried out a simulation using a data frame with the same noise (Gaussian) as our data frames. To this frame we have added randomly distributed galaxies that sample the CSB-magnitude plane. We have then used $Sextractor$ with the same detection parameters we have used above to try and detect our simulated galaxies. In Fig. 7 we distinguish in red those galaxies that are detected by $Sextractor$ and those that are not in blue. Ideally we would expect to detect all objects to the left of the dashed green curved line - an isophotal area of 800 pixels - and none to the right of this line. This is essentially true for the none detections, but the result is rather mixed for the detections. Some galaxies are confused with others and so are mis-measured, but clearly almost all large LSB galaxies are not detected by $Sextractor$ - we have indicated this region on Fig. 7.  We have also added a green box to Fig. 7, which defines the parameters of the LSB galaxies found by van Dokkum et al. (2015). Above we described how we miss galaxies with isophotal areas less than 800 pixels due to our chosen $Sextractor$ detection parameters and how further we miss galaxies with CSBs less than 22.5 $\mu g$ and scale sizes $\alpha < 3$ arc sec due to the necessity of separating cluster from background. We now find that if LSB galaxies become too large ($h > 10$ arc sec) we would also fail to find them. The large LSB galaxies are not detected because $Sextractor$ generally breaks them up into smaller components. Even though we switched the deblending option off, for such large LSB galaxies $Sextractor$ still tries to break them into smaller parts. We are in essence sampling an even smaller part of the CSB-magnitude plane than we first thought - we are now restricted to objects, along with our other criteria,  with scale sizes less than about 10 arc sec. However, the small region we are able to 'observe' is newly explored in the Virgo cluster and we have found new galaxies residing within it. Our simulation indicates that we are about 90\% complete for galaxies in this small region. This is all summarised in Fig. 7 where we have labeled areas of the CSB-magnitude plane\footnote{Sometimes referred to as the Bi-variate Brightness plane.}. Most important is the area of this plane we have investigated - indicated by the black dashed-line box. This box is but a small fraction of the  available parameter space that we already know galaxies exist within and an even smaller fraction of that which galaxies might occupy.

It is currently impossible to estimate the contribution to the cluster luminosity function of galaxies not detected because they lie in regions of the CSB-magnitude plane so far unobserved. It is clear from Fig. 1 and 2 that previous Virgo detections, Local Group galaxies and our newly detected galaxies do not lie on any well defined CSB magnitude relation that might help us predict the density of points across the CSB-magnitude plane. The only way forward is to 'observe' new regions of the CSB-magnitude plane and measure what is there. In the following section we will make a current best estimate of the luminosity function using all of the known galaxies.

\begin{figure}
\vspace{-2.0cm}
\centering
\includegraphics[scale=0.44]{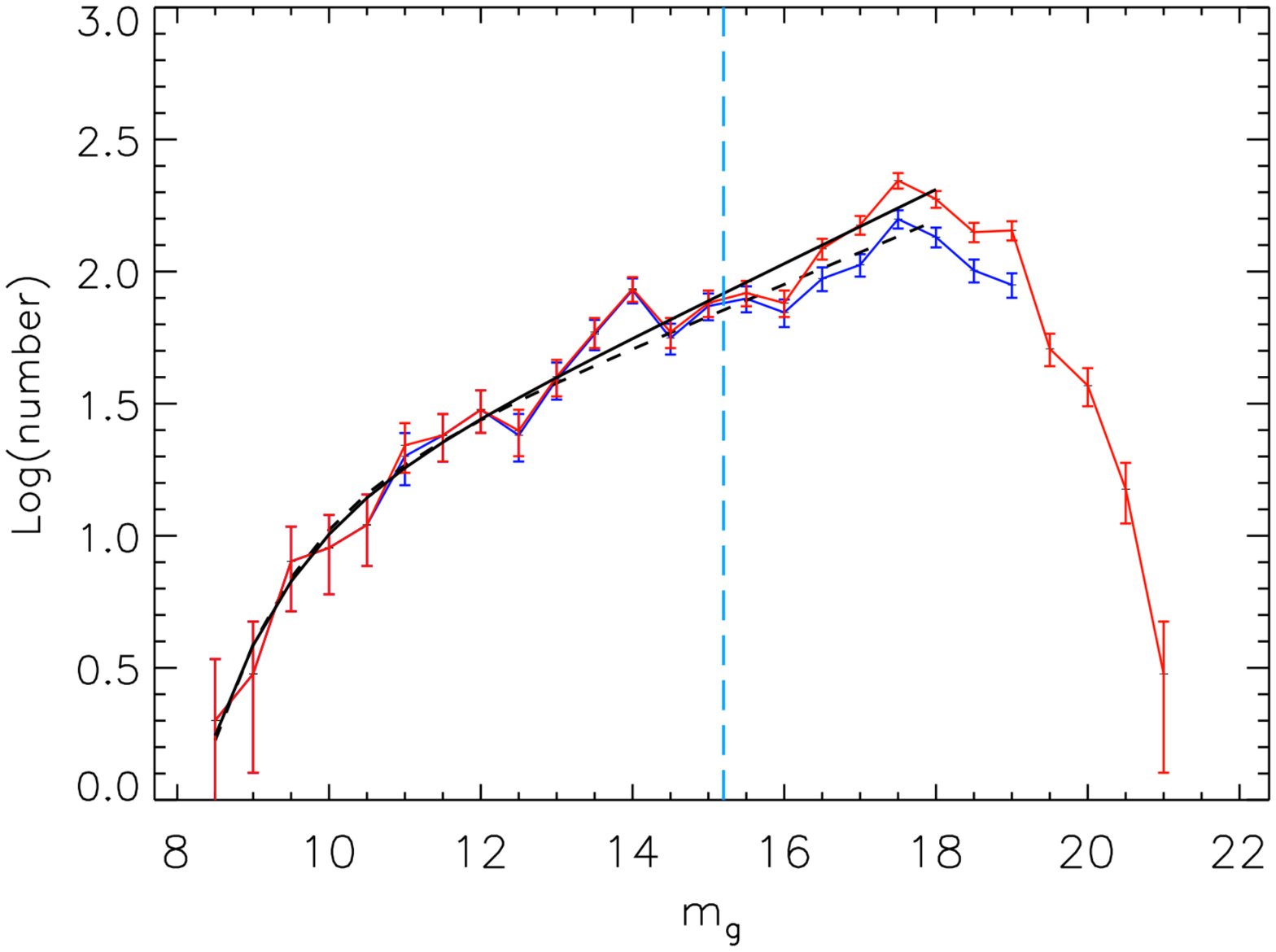}
\vspace{-3.5cm}
\caption{The Virgo cluster luminosity function. The dark blue line is the VCC data. The red line is the combined VVC, EVCC and LSBVCC data. In each case Poisson error bars are shown. The black dashed line is a Schechter function fit to the VCC data only. The black solid line is a Schechter function fit to the combined data. The vertical light blue dashed line marks the extent of the Montero-Dorta and Prada (2009) SDSS $g$ band luminosity function data too faint magnitudes given a Virgo distance modulus of 31.2. This line also marks the boundary between dwarf and brighter galaxies as defined in the introduction i.e. $M_{g}=-16.0$.}
\end{figure} 
 
\section{The Luminosity function}
To construct an up-to-date luminosity function we will combine data from the three surveys that have essentially covered the Virial radii of sub-clusters A and B. These are the photographic survey of Binggeli et al. (1985) (VCC), the rather shallow SDSS CCD survey of Kim et al. (2014) (EVCC) and the deeper NGVS CCD survey described here (LSBVCC). We will simply convert the observed apparent magnitudes into those applicable to the NGVS data using $g_{CFHT}=g_{SDSS}-0.09$ or $g_{CFHT}=B-0.61$ as described above. We have used a non-linear least squares fit to obtain Schechter function parameters of the luminosity function data. The given $1\sigma$ errors on the fitted parameters come from the Poisson errors on the individual data points and the goodness of fit to the data. The fit to the luminosity function data is carried out in the $8.5<g<18.0$ range as at $g>18.0$ the number count of galaxies drops dramatically, suggesting that we are incomplete at fainter magnitudes (Fig. 8).

The derived cluster luminosity function and best fitting Schechter function are shown in Fig. 8. The blue line is for the VCC galaxies with the red line being the total including VCC, EVCC and LSBVCC. The LSBVCC adds mainly to the faint end and clearly there are insufficient galaxies to maintain the VCC faint end slope beyond $g \approx 18.0$. We would require about a factor of twenty more galaxies at $g=20$ to maintain the current slope (this is from about 25 to 500 galaxies in total). Either we have missed a considerable number of galaxies or the luminosity function takes a sharp dive beyond $g \approx 18.0$ - we suspect the former. 

A Schechter luminosity function fit to the VCC data gives a faint end slope of $-1.30 \pm 0.03$ ($M^{*}=9.2 \pm 0.2$) while for the total data we find $-1.35 \pm 0.03$ ($M^{*}=9.1 \pm 0.2$). Our Schechter function fit only includes sources brightward of $g=18.0$, which is approximately the magnitude limit for the VCC and EVCC data. Including the EVCC and LSBVCC data leads to a negligible change in slope from that already measured using the VCC data. The Virgo cluster faint end slope is marginally steeper than that measured for field (all) galaxies by Montero-Dorta and Prada (2009) using SDSS $g$ band data - they measure  a faint end slope of $-1.10 \pm 0.03$. The Montero-Dorta and Prada (2009) data extends to $M_{g}=-16.0$ so, by going to fainter magnitudes and surface brightnesses we find that the faint end slope does become slightly steeper in this cluster environment. Whether this would be true for the field, if we could observe to these fainter limits, is still not clear or possibly the cluster environment plays a role in producing larger numbers of low luminosity galaxies (Tully et al. 2002). Currently we are not finding anywhere near sufficient Virgo cluster galaxies to steepen the luminosity function to the levels predicted by the numerical simulations.

\section{Conclusions}
The NGVS data offers the opportunity to investigate the LSB dwarf galaxy population of the Virgo cluster to surface brightness levels never before investigated. However, we are hampered by an inherent difficulty in separating cluster galaxies from background galaxies and a detection method that is not suited to the detection of extended LSB sources. The result is that currently we are only able to detect galaxies over the very limited range of $22.5 \le \mu^{g}_{0} \le 26.0$ $\mu g$ and $3.0 \le h \le 10.0$ arc sec, which leads to objects that will have $15.5 \le m_{g} \le 21.6 $ ($-15.7 \le M_{g} \le -9.6 $). Smaller scale sized galaxies are confused with the background, while larger galaxies are split into numerous smaller objects by the detection algorithm.

Given the limited area of the CSB-magnitude plane explored we have found 303 new Virgo cluster LSB dwarf galaxies. We are confident that these are genuine cluster members because we find no galaxies with these photometric properties on the 'background' fields and the surface density of these newly found galaxies falls with radius in just the same way as the VVC galaxies does.

Although we are sure that we are still missing large numbers of galaxies over the magnitude range $15.5 \le m_{g} \ge 21.6 $ we have put together the three large surveys that cover the Virial radius of the two sub-clusters A and B (VCC, EVCC and LSBVCC) to construct a current best estimate $g$ band luminosity function for the Virgo cluster. Adding galaxies from the EVCC and LSBVCC does not significantly alter the faint end slope derived using only the VCC data. The faint end slope is marginally steeper than that found for field galaxies. There is no evidence for a faint end slope steep enough to correspond with galaxy formation models, unless these models invoke either strong feedback processes or warm dark matter.

We plan to make further progress by investigating the use of colour information to separate foreground from background objects and pixel clustering algorithms to make our object detection more efficient.

\vspace{0.5cm}
\noindent
\large
{\bf Acknowledgements} \\
The observational data were obtained with MegaPrime/MegaCam, a joint project of CFHT and CEA/IRFU, at the Canada-France-Hawaii Telescope (CFHT) which is operated by the National Research Council (NRC) of Canada, the Institut National des Science de l'Univers of the Centre National de la Recherche Scientifique (CNRS) of France, and the University of Hawaii. 

This research used the facilities of the Canadian Astronomy Data Centre operated by the National Research Council of Canada with the support of the Canadian Space Agency.

\vspace{0.5cm}
\noindent
\large
{\bf References} \\
Abazajian K. et al., ApJS, 182, 543 \\
Bechtol et al., 2015, arXiv:1503.02584v1 \\
Benson A., Lacey C., Baugh C., Cole S., and Frenk C., 2002, MNRAS, 333, 156 \\
Binggeli B., et al., 1985, AJ, 90, 1681 \\
Blanton et al., 2003, ApJ, 592, 819 \\
Bohringer et al., 1994, Nature, 368, 828 \\
Boselli et al., 2011, AA, 528, 107 \\
Bothun G., Impey C., Malin D. and  Mould J., 1987, AJ, 94, 23 \\
Chung, A., van Gorkom J., Kenney J., Crowl H. and Vollmer B., AJ, 139, 1741
Cote P., et al., 2004, AJSS, 153, 223 \\
Davies J., et al., 2012, MNRAS, 419, 3505 \\
Davies J., et al., 2014, MNRAS, 438, 1922 \\
Davies J., Roberts S. and Sabatini S., 2005, MNRAS, 356, 794  \\
Driver S. et al., 2015, In `The Universe of Digital Sky Surveys', Naples, Italy, Nov. 2014 \\
Farrarese et al., 2012, ApJSS, 200, 4 \\
Gavazzi G., Boselli A., Scodeggio M., Pierini D. and Belsole E., 1999, MNRAS, 304, 595 \\
Gavazzi, G. Boselli, A. Donati, A. Franzetti, P. and Scodeggio, M., 2003, A\&A, 400, 451 \\
Giovanelli, R., et al., 2005, AJ, 130, 2598 \\
Impey C., Bothun G. and Malin D., ApJ, 330, 634 \\
James P., Knapen  J., Shane N., Baldry I. and de Jong R., 2008, AA, 482, 507 \\
Kauffman G., White, S., Guiderdoni B., 1993, MNRAS, 264, 201 \\
Kazantzidis S., Mayer L., Mastropietro C., Diemand J., Stadel J., Moore B., 2004, ApJ, 608, 663 \\
Kim et al., 2014, ApJS, 215, 22 \\
Klypin A., Kravtsov A., Valenzuela O., Prada F., 1999, ApJ, 522, 82 \\
Lovell, M., Frenk C., Eke V., Jenkins A., Gao L. and Theuns T., 2014, MNRA, 439, 300 \\
McConnachie A., 2012, AJ, 144, 4 \\
Mei S., et al., 2007, ApJ, 655, 144 \\
Merritt A., van Dokkum P., and Abraham R., 2014, ApJ, 787, 37 \\
Moore B. et al., 1999, ApJ, 524, L19 \\
Neugebauer G. et al., 1984, ApJ, 278, L1 \\
Norberg et al., 2002, MNRAS, 336, 907 \\
Power et al., 2015, In 'Advancing Astrophysics with the Square Kilometre Array', Giardini Naxos, Italy, June  2014 (arXiv150101564) \\
Sabatini S., et al., 2003, MNRAS, 341, 981 \\
Sabatini S., et al., 2005, MNRAS, 357, 819 \\
Shapley H. and Ames A., Harvard College Observatory Circular, vol. 294, pp.1-8 \\
Skrutskie M., et al., 2006, AJ, 131, 1163 \\
Somerville R., Primack J., and Faber S., 2001, MNRAS, 320, 504 \\
Stadel et al., 2009, MNRAS, 398, L21 \\
Stoehr F., White S., Tormen G., Springel V., 2002, MNRAS, 335, L8 \\
Taylor R., 2010, PhD thesis, Cardiff University, UK \\
Tully R., Somerville R., Trentham N. and Verheijen M., 2002, ApJ, 569, 573 \\
van Dokkum P., Abraham R., Merritt A., Zhang J., Geha M. and Conroy C., 2015, ApJ, 798, 45 \\
Warren S. et al., 2007, MNRAS, 375, 213 \\
Weisz D., et al., 2014, ApJ, 789, 147 \\

\setcounter{table}{0}
\onecolumn
\begin{center}
\begin{longtable}{cllccc} \hline
(1)  & (2)  & (3)  & (4) & (5) & (6)   \\ 
Name    &   RA        &     Dec   & $m_{g}$   & $\mu_{0}^{g}$ & $\alpha$  \\ 
             & (J2000)  &  (J2000) &               &                         & (arc sec) \\ \hline
LSBVCC001 &  12  31  31.8 &  11  36  10.8 &  17.8 &  23.0 &  4.4 \\
LSBVCC002 &  12  30  58.7 &  11  42  29.2 &  17.3 &  23.4 &  6.6 \\
LSBVCC003 &  12  33  46.9 &  11  46  55.6 &  18.1 &  24.0 &  5.8 \\
LSBVCC004 &  12  34  11.9 &  11  48  49.0 &  17.9 &  23.0 &  4.3 \\
LSBVCC005 &  12  31   3.9 &  11  50   7.8 &  17.4 &  23.3 &  6.0 \\
LSBVCC006 &  12  31  56.3 &  11  58  21.0 &  17.2 &  23.0 &  5.6 \\
LSBVCC007 &  12  33  52.5 &  12   7   3.7 &  20.3 &  24.8 &  3.2 \\
LSBVCC008 &  12  30  17.5 &  12  14  27.2 &  18.4 &  23.7 &  4.5 \\
LSBVCC009 &  12  33  41.0 &  12  22  57.4 &  18.5 &  23.6 &  4.1 \\
LSBVCC010 &  12  32   7.6 &  12  26   5.6 &  17.6 &  23.2 &  5.2 \\
LSBVCC011 &  12  33  40.8 &  12  34  16.0 &  17.7 &  23.3 &  5.2 \\
LSBVCC012 &  12  31  44.0 &  12  36  44.6 &  18.5 &  23.4 &  3.8 \\
LSBVCC013 &  12  32  34.7 &  12  38  21.1 &  18.0 &  24.7 &  8.6 \\
LSBVCC014 &  12  34   1.5 &  12  43  12.7 &  19.0 &  24.0 &  3.9 \\
LSBVCC015 &  12  32  55.4 &  12  45  34.2 &  20.5 &  25.0 &  3.2 \\
LSBVCC016 &  12  33  51.1 &  12  57  30.6 &  19.1 &  23.7 &  3.3 \\
LSBVCC017 &  12  31  36.4 &  13   5  19.3 &  19.0 &  24.1 &  4.3 \\
LSBVCC018 &  12  31  10.5 &  13   5  51.4 &  17.9 &  23.0 &  4.2 \\
LSBVCC019 &  12  31  12.7 &  13   7  27.5 &  18.0 &  22.6 &  3.4 \\
LSBVCC020 &  12  33   6.4 &  13  18  10.8 &  18.4 &  24.1 &  5.6 \\
LSBVCC021 &  12  30  24.4 &  13  19  57.4 &  18.0 &  23.7 &  5.6 \\
LSBVCC022 &  12  30  55.7 &  13  20  52.8 &  17.8 &  23.2 &  4.9 \\
LSBVCC023 &  12  31  24.5 &  13  20  56.0 &  20.2 &  24.7 &  3.2 \\
LSBVCC024 &  12  33  25.3 &  13  24  59.0 &  18.9 &  24.8 &  5.9 \\
LSBVCC025 &  12  30  49.5 &  13  30  18.4 &  19.1 &  24.7 &  5.2 \\
LSBVCC026 &  12  32  50.3 &  13  29  51.4 &  20.0 &  24.5 &  3.2 \\
LSBVCC027 &  12  31  10.1 &  13  30  42.8 &  19.6 &  24.3 &  3.5 \\
LSBVCC028 &  12  33   5.0 &  13  32  35.2 &  18.3 &  23.1 &  3.7 \\
LSBVCC029 &  12  30  57.6 &  13  37  11.3 &  18.1 &  23.2 &  4.1 \\
LSBVCC030 &  12  33  19.4 &  13  37  12.7 &  20.6 &  25.0 &  3.0 \\
LSBVCC031 &  12  31  39.7 &  13  38  53.5 &  19.3 &  24.3 &  3.9 \\
LSBVCC032 &  12  31  35.7 &  13  49  27.1 &  17.9 &  24.4 &  7.9 \\
LSBVCC033 &  12  32  31.8 &  13  51   2.9 &  18.2 &  22.7 &  3.1 \\
LSBVCC034 &  12  30  42.6 &  13  52  51.6 &  19.7 &  24.2 &  3.2 \\
LSBVCC035 &  12  33  17.7 &  13  56   6.0 &  17.4 &  23.0 &  5.1 \\
LSBVCC036 &  12  31  36.2 &  14   4  17.8 &  17.7 &  23.4 &  5.3 \\
LSBVCC037 &  12  32  15.9 &  14  11  17.9 &  20.0 &  24.4 &  3.0 \\
LSBVCC038 &  12  33   9.0 &  10  50  10.3 &  17.8 &  22.6 &  3.6 \\
LSBVCC039 &  12  30  53.7 &  10  54  42.1 &  19.7 &  24.1 &  3.0 \\
LSBVCC040 &  12  32  44.9 &  10  56  56.8 &  18.0 &  23.5 &  5.0 \\
LSBVCC041 &  12  31  36.9 &  11   0  27.4 &  17.4 &  22.7 &  4.8 \\
LSBVCC042 &  12  32   0.2 &  11   1  24.2 &  18.0 &  23.0 &  4.1 \\
LSBVCC043 &  12  31  44.7 &  11  13  58.4 &  19.4 &  24.3 &  3.8 \\
LSBVCC044 &  12  33  15.0 &  11  13  50.9 &  19.0 &  23.8 &  3.6 \\
LSBVCC045 &  12  32  51.2 &  11  17  44.5 &  18.9 &  23.8 &  3.7 \\
LSBVCC046 &  12  31  60.0 &  11  18   5.8 &  19.2 &  24.5 &  4.6 \\
LSBVCC047 &  12  30  25.9 &  11  26  13.6 &  18.4 &  23.7 &  4.5 \\
LSBVCC048 &  12  32  20.0 &   9  46  20.3 &  19.2 &  24.5 &  4.7 \\
LSBVCC049 &  12  33   5.1 &   9  59  10.7 &  18.1 &  22.9 &  3.6 \\
LSBVCC050 &  12  31  30.7 &  10   0  19.8 &  18.6 &  23.8 &  4.4 \\
LSBVCC051 &  12  30  48.1 &  10   8  22.6 &  18.8 &  24.2 &  4.8 \\
LSBVCC052 &  12  33  53.2 &  10  24  51.1 &  19.7 &  24.6 &  3.9 \\
LSBVCC053 &  12  35  34.7 &  11  37  13.4 &  18.5 &  24.5 &  6.4 \\
LSBVCC054 &  12  34  30.3 &  11  44   4.2 &  16.9 &  23.1 &  7.2 \\
LSBVCC055 &  12  34   6.5 &  11  50  13.9 &  17.9 &  23.1 &  4.3 \\
LSBVCC056 &  12  36  27.1 &  11  53  20.4 &  17.6 &  23.9 &  7.2 \\
LSBVCC057 &  12  35   6.6 &  11  54   2.5 &  17.6 &  23.8 &  6.9 \\
LSBVCC058 &  12  35  41.7 &  12  12  27.7 &  19.6 &  24.8 &  4.2 \\
LSBVCC059 &  12  35  40.8 &  12  14   7.1 &  19.7 &  24.3 &  3.3 \\
LSBVCC060 &  12  34  36.8 &  12  25  52.0 &  20.9 &  25.5 &  3.4 \\
LSBVCC061 &  12  36  32.6 &  12  31   4.1 &  18.9 &  24.1 &  4.4 \\
LSBVCC062 &  12  35  26.8 &  12  31  41.5 &  18.5 &  23.2 &  3.5 \\
LSBVCC063 &  12  35  44.1 &  12  36  34.2 &  18.7 &  24.6 &  6.0 \\
LSBVCC064 &  12  37  29.0 &  12  52  30.4 &  20.2 &  24.6 &  3.0 \\
LSBVCC065 &  12  37  37.6 &  12  54  35.6 &  19.1 &  23.5 &  3.1 \\
LSBVCC066 &  12  36  52.5 &  13  17   2.0 &  18.9 &  23.7 &  3.6 \\
LSBVCC067 &  12  36  25.3 &  13  29  28.0 &  19.4 &  24.2 &  3.8 \\
LSBVCC068 &  12  37  27.7 &  13  47  32.3 &  19.4 &  24.8 &  4.9 \\
LSBVCC069 &  12  35  43.5 &  14   4  44.4 &  20.1 &  25.4 &  4.7 \\
LSBVCC070 &  12  34  42.4 &  14  13  22.8 &  17.1 &  23.2 &  6.5 \\
LSBVCC071 &  12  34  46.7 &  14  17  32.3 &  19.1 &  23.6 &  3.1 \\
LSBVCC072 &  12  37  31.6 &  11   2  13.2 &  19.8 &  24.7 &  3.8 \\
LSBVCC073 &  12  34   6.0 &  11   3  15.8 &  19.3 &  24.1 &  3.6 \\
LSBVCC074 &  12  36  37.4 &  11   9  12.2 &  17.3 &  23.9 &  8.4 \\
LSBVCC075 &  12  37  55.0 &  11   8  56.0 &  18.6 &  23.9 &  4.7 \\
LSBVCC076 &  12  34  15.5 &  11  28   3.0 &  18.8 &  24.8 &  6.4 \\
LSBVCC077 &  12  37   3.2 &  11  28  42.6 &  19.2 &  24.4 &  4.3 \\
LSBVCC078 &  12  34  57.1 &  10  3 4  5.7 &  18.6 &  23.3 &  3.5 \\
LSBVCC079 &  12  38  40.8 &  11  58  42.6 &  19.5 &  24.8 &  4.6 \\
LSBVCC080 &  12  41  14.0 &  12  14  56.8 &  20.0 &  24.4 &  3.1 \\
LSBVCC081 &  12  41  28.1 &  12  25  40.8 &  18.2 &  22.8 &  3.3 \\
LSBVCC082 &  12  38  12.1 &  10  42   6.8 &  18.7 &  23.9 &  4.3 \\
LSBVCC083 &  12  40  55.6 &  10  42  10.8 &  19.9 &  24.8 &  3.8 \\
LSBVCC084 &  12  40  18.3 &  10  59  48.1 &  18.1 &  23.0 &  3.9 \\
LSBVCC085 &  12  38  17.1 &  11  10  50.5 &  18.7 &  24.2 &  5.1 \\
LSBVCC086 &  12  39  31.7 &  11  27  15.5 &  18.5 &  25.2 &  8.5 \\
LSBVCC087 &  12  40   4.8 &  11  34  16.0 &  18.6 &  23.3 &  3.5 \\
LSBVCC088 &  12  27  33.2 &  11  31  54.1 &  19.5 &  24.2 &  3.5 \\
LSBVCC089 &  12  29  39.1 &  11  37  58.4 &  17.3 &  24.1 &  8.9 \\
LSBVCC090 &  12  28  59.5 &  11  55  25.0 &  20.1 &  25.0 &  3.7 \\
LSBVCC091 &  12  27  43.4 &  11  58   4.4 &  19.9 &  24.7 &  3.7 \\
LSBVCC092 &  12  29  41.3 &  12   2  45.2 &  18.2 &  23.9 &  5.4 \\
LSBVCC093 &  12  28  59.1 &  12   2  30.1 &  20.1 &  24.6 &  3.1 \\
LSBVCC094 &  12  28  49.2 &  12   7  56.3 &  18.2 &  23.6 &  4.7 \\
LSBVCC095 &  12  29   5.2 &  12   9  15.1 &  19.4 &  24.4 &  4.1 \\
LSBVCC096 &  12  27  19.5 &  12  13  13.8 &  19.6 &  24.7 &  4.0 \\
LSBVCC097 &  12  27  29.7 &  12  15   6.8 &  18.7 &  23.6 &  3.8 \\
LSBVCC098 &  12  28  26.6 &  12  20  42.0 &  18.9 &  24.2 &  4.5 \\
LSBVCC099 &  12  27  16.7 &  10  41  55.7 &  19.5 &  24.0 &  3.2 \\
LSBVCC100 &  12  26  56.5 &  10  47  53.2 &  20.6 &  25.1 &  3.1 \\
LSBVCC101 &  12  29  59.4 &  10  51  22.7 &  17.8 &  23.6 &  6.0 \\
LSBVCC102 &  12  28  49.7 &  10  55  14.9 &  18.9 &  24.3 &  4.8 \\
LSBVCC103 &  12  29  21.1 &  10  59  39.1 &  19.8 &  24.8 &  4.1 \\
LSBVCC104 &  12  29  12.9 &  10  59  26.5 &  19.5 &  24.1 &  3.2 \\
LSBVCC105 &  12  27  44.3 &  11  12  52.9 &  17.1 &  22.6 &  5.0 \\
LSBVCC106 &  12  26  34.0 &  11  17   6.0 &  17.9 &  23.9 &  6.4 \\
LSBVCC107 &  12  29  57.1 &  11  24   3.6 &  19.4 &  24.9 &  4.8 \\
LSBVCC108 &  12  28  23.4 &  11  34  46.6 &  18.0 &  22.7 &  3.5 \\
LSBVCC109 &  12  40   1.8 &  13  39  16.6 &  18.3 &  23.6 &  4.4 \\
LSBVCC110 &  12  40  15.4 &  13  52  44.0 &  20.3 &  24.7 &  3.1 \\
LSBVCC111 &  12  38   0.5 &  14  10  39.4 &  17.6 &  22.5 &  3.8 \\
LSBVCC112 &  12  38  44.5 &  14  39  47.5 &  18.6 &  23.0 &  3.1 \\
LSBVCC113 &  12  41  10.6 &  15   2  55.0 &  17.6 &  23.9 &  7.1 \\
LSBVCC114 &  12  40  58.7 &  15  10   6.6 &  20.6 &  25.0 &  3.1 \\
LSBVCC115 &  12  41  13.9 &  15  22  10.6 &  17.5 &  24.2 &  8.8 \\
LSBVCC116 &  12  41  41.8 &  16  13  22.4 &  18.2 &  23.8 &  5.2 \\
LSBVCC117 &  12  42  37.8 &  13  36  22.7 &  17.3 &  22.6 &  4.5 \\
LSBVCC118 &  12  42  56.5 &  13  45  55.4 &  19.5 &  23.9 &  3.0 \\
LSBVCC119 &  12  42  36.2 &  16  18  23.4 &  17.1 &  22.9 &  5.6 \\
LSBVCC120 &  12  43  13.0 &  16  20  36.2 &  17.9 &  23.6 &  5.4 \\
LSBVCC121 &  12  41  49.1 &   8  18   2.9 &  18.1 &  23.6 &  5.2 \\
LSBVCC122 &  12  47  12.4 &  10  35  20.8 &  20.0 &  25.4 &  4.9 \\
LSBVCC123 &  12  47  48.8 &  10  38  52.1 &  17.9 &  22.7 &  3.6 \\
LSBVCC124 &  12  45  54.1 &  10  39  31.3 &  19.4 &  23.9 &  3.2 \\
LSBVCC125 &  12  46  36.2 &  10  47   4.6 &  18.1 &  24.2 &  6.5 \\
LSBVCC126 &  12  46  18.1 &  11  11   7.4 &  18.7 &  24.0 &  4.6 \\
LSBVCC127 &  12  49   4.4 &  11  10  39.0 &  20.0 &  24.8 &  3.8 \\
LSBVCC128 &  12  48   0.2 &  11  22  21.7 &  19.4 &  24.4 &  3.9 \\
LSBVCC129 &  12  38   2.2 &  14  24  53.6 &  20.2 &  24.8 &  3.0 \\
LSBVCC130 &  12  37  52.0 &  14  28   1.9 &  17.7 &  23.5 &  6.0 \\
LSBVCC131 &  12  34  18.6 &  14  28  22.1 &  19.2 &  23.6 &  3.0 \\
LSBVCC132 &  12  34  54.7 &  14  39   1.8 &  18.9 &  24.1 &  4.0 \\
LSBVCC133 &  12  37  27.1 &  15   8  57.5 &  18.8 &  24.2 &  4.0 \\
LSBVCC134 &  12  35  41.7 &  15   8  30.1 &  19.3 &  24.5 &  4.0 \\
LSBVCC135 &  12  34  25.7 &  15  26  42.7 &  19.5 &  23.9 &  3.0 \\
LSBVCC136 &  12  36  42.4 &  15  28  28.2 &  18.8 &  23.1 &  3.0 \\
LSBVCC137 &  12  37  50.2 &  15  43  39.0 &  17.4 &  23.4 &  6.0 \\
LSBVCC138 &  12  34  45.2 &  16   1  26.8 &  17.7 &  23.0 &  4.0 \\
LSBVCC139 &  12  38   8.2 &  16   4  18.1 &  17.2 &  23.2 &  6.0 \\
LSBVCC140 &  12  37  46.6 &  16  10  31.8 &  18.0 &  22.6 &  3.0 \\
LSBVCC141 &  12  35   3.5 &  16  31  59.2 &  19.7 &  24.2 &  3.0 \\
LSBVCC142 &  12  34   9.8 &  16  42  42.8 &  17.9 &  23.6 &  5.0 \\
LSBVCC143 &  12  35  24.2 &   6  30   4.3 &  18.4 &  23.6 &  4.0 \\
LSBVCC144 &  12  41  39.3 &   9  12  31.3 &  17.8 &  24.6 &  9.0 \\
LSBVCC145 &  12  39  42.8 &   9  18  34.6 &  19.4 &  24.3 &  3.0 \\
LSBVCC146 &  12  39  35.9 &   7   9  48.2 &  20.5 &  25.5 &  4.0 \\
LSBVCC147 &  12  40  28.1 &   7   9  56.2 &  20.8 &  25.5 &  3.0 \\
LSBVCC148 &  12  39  48.5 &   7  18  51.5 &  19.0 &  24.1 &  4.0 \\
LSBVCC149 &  12  37  57.7 &   6  32  31.9 &  18.2 &  23.3 &  4.0 \\
LSBVCC150 &  12  38  42.7 &   6  40   7.3 &  18.0 &  22.6 &  3.0 \\
LSBVCC151 &  12  27  15.7 &  13  26  59.6 &  19.5 &  24.1 &  3.3 \\
LSBVCC152 &  12  26  47.6 &  13  40  49.4 &  18.5 &  22.9 &  3.0 \\
LSBVCC153 &  12  30  11.8 &  13  41  28.0 &  18.5 &  22.9 &  3.0 \\
LSBVCC154 &  12  29  54.2 &  13  57  23.0 &  18.6 &  23.3 &  3.5 \\
LSBVCC155 &  12  30  17.5 &  14   7  42.6 &  18.2 &  23.0 &  3.7 \\
LSBVCC156 &  12  28  27.2 &  14   9  20.2 &  16.7 &  22.7 &  6.4 \\
LSBVCC157 &  12  28  18.9 &  14   8  15.7 &  19.8 &  24.5 &  3.4 \\
LSBVCC158 &  12  28  49.6 &  14   9  28.1 &  17.6 &  23.4 &  5.8 \\
LSBVCC159 &  12  29  29.4 &  14  10  10.9 &  18.6 &  23.9 &  4.7 \\
LSBVCC160 &  12  28  32.7 &  14  11  28.3 &  20.3 &  24.8 &  3.1 \\
LSBVCC161 &  12  29   9.8 &  14  13  40.1 &  20.0 &  24.7 &  3.6 \\
LSBVCC162 &  12  28  52.7 &  14  23  39.1 &  18.8 &  23.2 &  3.1 \\
LSBVCC163 &  12  27  57.4 &  14  28  14.5 &  18.3 &  24.4 &  6.5 \\
LSBVCC164 &  12  29  57.6 &  14  28  21.0 &  18.7 &  23.7 &  4.0 \\
LSBVCC165 &  12  28  44.7 &  14  29  38.0 &  19.5 &  24.0 &  3.0 \\
LSBVCC166 &  12  27   6.1 &  14  55  15.6 &  19.8 &  24.8 &  3.9 \\
LSBVCC167 &  12  23  38.5 &  12  37  40.1 &  18.0 &  22.6 &  3.3 \\
LSBVCC168 &  12  25  23.6 &  12  40  31.1 &  19.2 &  24.2 &  4.0 \\
LSBVCC169 &  12  25  26.6 &  12  40  36.8 &  18.6 &  23.3 &  3.5 \\
LSBVCC170 &  12  23  53.0 &  12  46  21.4 &  18.0 &  23.9 &  5.9 \\
LSBVCC171 &  12  25  49.1 &  12  48  15.8 &  18.6 &  23.1 &  3.2 \\
LSBVCC172 &  12  24  18.7 &  12  54  45.4 &  18.5 &  23.1 &  3.3 \\
LSBVCC173 &  12  25  30.3 &  12  58  37.2 &  18.9 &  24.6 &  5.6 \\
LSBVCC174 &  12  26   4.2 &  12  59  16.4 &  20.0 &  24.8 &  3.7 \\
LSBVCC175 &  12  25  13.2 &  13   1  31.4 &  17.4 &  23.1 &  5.5 \\
LSBVCC176 &  12  25  49.1 &  13  16  59.5 &  19.8 &  24.6 &  3.7 \\
LSBVCC177 &  12  24  45.5 &  13  20  25.1 &  17.8 &  22.7 &  3.7 \\
LSBVCC178 &  12  24  11.2 &  13  22  24.6 &  18.4 &  23.6 &  4.3 \\
LSBVCC179 &  12  22  47.7 &  13  21  11.9 &  20.3 &  24.7 &  3.1 \\
LSBVCC180 &  12  24  34.3 &  13  22  24.6 &  19.4 &  23.8 &  3.1 \\
LSBVCC181 &  12  23  11.6 &  13  25   7.7 &  18.8 &  23.8 &  4.0 \\
LSBVCC182 &  12  24  21.8 &  13  25   6.6 &  18.3 &  23.2 &  3.8 \\
LSBVCC183 &  12  25  48.0 &  13  51  16.2 &  18.3 &  23.2 &  3.7 \\
LSBVCC184 &  12  25  31.7 &  14   9   7.6 &  16.9 &  22.5 &  5.4 \\
LSBVCC185 &  12  25  41.1 &  14  11  31.2 &  19.3 &  24.7 &  4.8 \\
LSBVCC186 &  12  24  12.7 &  14  29  37.0 &  18.8 &  23.5 &  3.4 \\
LSBVCC187 &  12  25  55.1 &  14  38  29.0 &  19.2 &  23.8 &  3.3 \\
LSBVCC188 &  12  24  53.1 &  14  39  18.4 &  18.2 &  22.8 &  3.4 \\
LSBVCC189 &  12  25   0.4 &  15   5  37.3 &  19.3 &  23.8 &  3.1 \\
LSBVCC190 &  12  24  35.1 &  15   9  51.1 &  18.8 &  24.0 &  4.2 \\
LSBVCC191 &  12  24  51.1 &  15  23  40.9 &  19.6 &  24.2 &  3.4 \\
LSBVCC192 &  12  22  24.3 &  15  28  16.0 &  17.1 &  23.1 &  6.2 \\
LSBVCC193 &  12  22  55.5 &  15  33  34.2 &  18.8 &  23.7 &  3.7 \\
LSBVCC194 &  12  25  38.8 &  15  48  52.2 &  19.7 &  25.5 &  5.5 \\
LSBVCC195 &  12  22  12.9 &  12  41  47.4 &  18.3 &  23.7 &  4.9 \\
LSBVCC196 &  12  21  29.3 &  12  47  53.5 &  19.0 &  23.4 &  3.0 \\
LSBVCC197 &  12  19  12.5 &  12  51   6.5 &  16.7 &  22.5 &  5.9 \\
LSBVCC198 &  12  20  36.4 &  12  53   6.0 &  18.3 &  23.5 &  4.4 \\
LSBVCC199 &  12  22  28.2 &  12  53  56.0 &  18.5 &  23.2 &  3.5 \\
LSBVCC200 &  12  18  42.5 &  12  56  42.4 &  20.2 &  25.2 &  3.9 \\
LSBVCC201 &  12  22  28.0 &  13   9  46.4 &  19.3 &  23.9 &  3.3 \\
LSBVCC202 &  12  21  15.1 &  13  20  46.3 &  16.5 &  22.5 &  6.4 \\
LSBVCC203 &  12  17  33.8 &  13  23  13.6 &  20.3 &  24.8 &  3.2 \\
LSBVCC204 &  12  16   8.5 &  13  27  45.4 &  19.0 &  23.8 &  3.6 \\
LSBVCC205 &  12  32  29.9 &  14  39  45.7 &  18.1 &  24.0 &  6.2 \\
LSBVCC206 &  12  32   3.9 &  15   3  17.6 &  18.6 &  23.5 &  3.9 \\
LSBVCC207 &  12  33  29.6 &  15  13  58.4 &  18.4 &  24.3 &  6.1 \\
LSBVCC208 &  12  44   6.6 &  12  41   2.8 &  19.5 &  23.9 &  3.0 \\
LSBVCC209 &  12  30   7.0 &   6  20  31.2 &  19.5 &  24.6 &  4.2 \\
LSBVCC210 &  12  25  39.6 &  16  16  58.4 &  18.9 &  24.5 &  5.3 \\
LSBVCC211 &  12  25  35.8 &  16  35  45.2 &  17.6 &  23.5 &  6.2 \\
LSBVCC212 &  12  23  25.9 &  16  44  15.4 &  20.2 &  24.9 &  3.5 \\
LSBVCC213 &  12  22  24.1 &  17   1  12.0 &  16.4 &  22.7 &  7.2 \\
LSBVCC214 &  12  24  15.2 &  17   7  40.8 &  18.5 &  23.1 &  3.4 \\
LSBVCC215 &  12  23  56.4 &   6   1  52.7 &  18.8 &  23.2 &  3.0 \\
LSBVCC216 &  12  26  12.3 &   6   5  10.7 &  19.0 &  24.4 &  4.7 \\
LSBVCC217 &  12  25  11.1 &   6  18  49.7 &  18.2 &  23.0 &  3.6 \\
LSBVCC218 &  12  24  25.8 &   4  56  31.6 &  20.1 &  24.5 &  3.0 \\
LSBVCC219 &  12  23   9.0 &   5  12  25.9 &  20.2 &  25.1 &  3.7 \\
LSBVCC220 &  12  25  37.0 &   5  20  27.2 &  17.8 &  23.0 &  4.4 \\
LSBVCC221 &  12  22  44.6 &   5  24  56.5 &  18.5 &  23.4 &  3.8 \\
LSBVCC222 &  12  26  14.9 &   5  26  15.0 &  19.2 &  23.6 &  3.1 \\
LSBVCC223 &  12  23  30.4 &   5  29  44.2 &  18.2 &  24.0 &  5.6 \\
LSBVCC224 &  12  23  16.9 &   5  49  47.3 &  21.1 &  25.6 &  3.1 \\
LSBVCC225 &  12  22  27.1 &  10   9  58.3 &  20.3 &  24.9 &  3.3 \\
LSBVCC226 &  12  20  17.5 &  10  19  13.4 &  17.9 &  23.9 &  6.4 \\
LSBVCC227 &  12  19  51.4 &  10  18  25.6 &  19.8 &  24.6 &  3.5 \\
LSBVCC228 &  12  21  29.0 &  10  29   9.2 &  17.8 &  22.9 &  4.1 \\
LSBVCC229 &  12  18  48.4 &   8  59  18.2 &  18.4 &  22.9 &  3.2 \\
LSBVCC230 &  12  20  44.9 &   9  22  24.2 &  19.6 &  23.9 &  3.0 \\
LSBVCC231 &  12  20  35.8 &   8  12   5.8 &  18.0 &  23.6 &  5.3 \\
LSBVCC232 &  12  21  55.8 &   8  20  45.2 &  18.6 &  23.5 &  3.7 \\
LSBVCC233 &  12  21  50.4 &   8  32  30.5 &  17.2 &  22.8 &  5.2 \\
LSBVCC234 &  12  18  57.7 &   8  38  39.8 &  18.6 &  24.1 &  5.1 \\
LSBVCC235 &  12  22  27.6 &   7   5  52.1 &  18.3 &  23.4 &  4.2 \\
LSBVCC236 &  12  40   1.0 &   9  48  53.6 &  19.5 &  24.2 &  3.4 \\
LSBVCC237 &  12  40   1.8 &   9  57  24.8 &  19.4 &  24.6 &  4.4 \\
LSBVCC238 &  12  40  27.1 &  10   7  42.2 &  18.4 &  24.0 &  5.2 \\
LSBVCC239 &  12  43  44.9 &  10   9  46.4 &  19.3 &  24.6 &  4.4 \\
LSBVCC240 &  12  45   9.7 &  10  37  41.5 &  17.6 &  23.5 &  6.0 \\
LSBVCC241 &  12  42  16.7 &   8  49  18.1 &  20.0 &  25.0 &  4.0 \\
LSBVCC242 &  12  44  38.7 &   9  23  36.2 &  20.1 &  25.0 &  3.7 \\
LSBVCC243 &  12  17  24.7 &  13  30  38.9 &  19.1 &  24.2 &  4.2 \\
LSBVCC244 &  12  15  56.5 &  13  39  22.7 &  20.1 &  24.5 &  3.0 \\
LSBVCC245 &  12  17  26.6 &  13  44  13.6 &  16.8 &  23.0 &  6.8 \\
LSBVCC246 &  12  14  50.8 &  13  59   0.2 &  18.3 &  22.8 &  3.1 \\
LSBVCC247 &  12  18  44.0 &  10  32  54.2 &  19.9 &  24.6 &  3.5 \\
LSBVCC248 &  12  17   7.6 &   8  44  28.0 &  18.1 &  22.6 &  3.2 \\
LSBVCC249 &  12  15  56.4 &   9  39  1.8  &  17.5 &  23.5 &  6.5 \\
LSBVCC250 &  12  15  32.9 &   9  39  40.0 &  19.0 &  23.7 &  3.4 \\
LSBVCC251 &  12  48  43.7 &  11  58   9.1 &  16.8 &  23.2 &  7.8 \\
LSBVCC252 &  12  48  25.2 &  12  14  12.5 &  18.7 &  23.5 &  3.5 \\
LSBVCC253 &  12  47  47.5 &  12  41  42.0 &  19.7 &  24.4 &  3.5 \\
LSBVCC254 &  12  46  31.4 &  13  19  32.2 &  19.5 &  24.5 &  4.1 \\
LSBVCC255 &  12  48   6.6 &  14  13  52.0 &  19.9 &  24.7 &  3.6 \\
LSBVCC256 &  12  49  25.6 &  14  54  51.8 &  18.9 &  24.1 &  4.4 \\
LSBVCC257 &  12  49  15.3 &  15  29  32.3 &  18.9 &  24.3 &  4.8 \\
LSBVCC258 &  12  48  32.2 &  15  44  22.2 &  18.0 &  23.2 &  4.5 \\
LSBVCC259 &  12  46  37.0 &  15  45  25.2 &  19.9 &  24.5 &  3.4 \\
LSBVCC260 &  12  46  13.2 &  10  14  50.6 &  18.7 &  23.4 &  3.4 \\
LSBVCC261 &  12  46  29.2 &  10  16  53.0 &  19.2 &  24.4 &  4.5 \\
LSBVCC262 &  12  47  45.8 &   9  13  48.7 &  17.8 &  23.3 &  5.0 \\
LSBVCC263 &  12  47   8.8 &   9  18  49.3 &  18.9 &  24.2 &  4.4 \\
LSBVCC264 &  12  47  27.1 &   9  25  44.8 &  18.5 &  24.2 &  5.5 \\
LSBVCC265 &  12  52  54.9 &  11  30  20.5 &  19.1 &  24.1 &  3.9 \\
LSBVCC266 &  12  50  45.4 &  11  54  44.6 &  19.5 &  24.0 &  3.3 \\
LSBVCC267 &  12  52   4.0 &  12  54  26.3 &  18.3 &  23.8 &  5.0 \\
LSBVCC268 &  12  50  54.4 &  10  46  11.6 &  19.1 &  23.9 &  3.8 \\
LSBVCC269 &  12  50  45.2 &  10  47  47.8 &  19.9 &  24.5 &  3.3 \\
LSBVCC270 &  12  52   5.9 &  11   1  56.6 &  18.2 &  23.5 &  4.4 \\
LSBVCC271 &  12  52  39.6 &  11   6  40.3 &  19.0 &  23.8 &  3.6 \\
LSBVCC272 &  12  51  11.0 &  11  14  33.4 &  17.5 &  23.3 &  5.9 \\
LSBVCC273 &  12  51   1.1 &  11  29   4.6 &  19.4 &  25.0 &  5.3 \\
LSBVCC274 &  12  26  28.6 &   8  47  42.4 &  18.7 &  23.7 &  4.1 \\
LSBVCC275 &  12  29  27.7 &   8  56  24.4 &  19.1 &  23.4 &  3.0 \\
LSBVCC276 &  12  26  34.4 &   8  59  34.1 &  18.5 &  23.7 &  4.4 \\
LSBVCC277 &  12  30   2.9 &   9  14  22.9 &  20.1 &  24.8 &  3.5 \\
LSBVCC278 &  12  29   8.1 &   9  26  37.0 &  16.5 &  22.8 &  7.3 \\
LSBVCC279 &  12  29  39.7 &   9  27  49.0 &  17.9 &  22.5 &  3.3 \\
LSBVCC280 &  12  30   2.0 &   9  28  26.4 &  18.3 &  22.9 &  3.3 \\
LSBVCC281 &  12  29  38.1 &   9  31  15.2 &  17.0 &  22.7 &  5.4 \\
LSBVCC282 &  12  25  55.5 &  10  43  16.0 &  18.6 &  24.1 &  4.9 \\
LSBVCC283 &  12  24   8.4 &  11   8  46.3 &  19.4 &  24.5 &  4.1 \\
LSBVCC284 &  12  23  31.9 &  11  16  14.9 &  18.0 &  23.4 &  4.6 \\
LSBVCC285 &  12  24  12.5 &  11  31  45.8 &  17.5 &  22.5 &  4.0 \\
LSBVCC286 &  12  24  56.5 &   9  55  36.5 &  20.7 &  25.1 &  3.0 \\
LSBVCC287 &  12  23  29.0 &  10   1  21.4 &  20.0 &  24.5 &  3.1 \\
LSBVCC288 &  12  23  17.5 &  10   3  24.5 &  20.5 &  25.0 &  3.1 \\
LSBVCC289 &  12  25  47.3 &  10   5  33.7 &  17.6 &  23.6 &  6.5 \\
LSBVCC290 &  12  23  24.5 &  10   9  24.8 &  20.1 &  24.7 &  3.4 \\
LSBVCC291 &  12  24   7.7 &  10  24  50.8 &  18.3 &  22.6 &  3.0 \\
LSBVCC292 &  12  25  27.3 &  10  25  10.9 &  21.0 &  25.5 &  3.2 \\
LSBVCC293 &  12  25  10.1 &  10  27  24.1 &  18.9 &  23.4 &  3.1 \\
LSBVCC294 &  12  24  30.9 &  10  31  32.9 &  20.2 &  24.7 &  3.1 \\
LSBVCC295 &  12  26  24.8 &  10  34  54.1 &  17.5 &  23.3 &  5.6 \\
LSBVCC296 &  12  23   6.5 &  10  35  24.4 &  20.3 &  24.7 &  3.0 \\
LSBVCC297 &  12  25  47.1 &  11  39  44.3 &  17.9 &  23.5 &  5.1 \\
LSBVCC298 &  12  25  55.7 &  11  48   3.6 &  19.2 &  24.4 &  4.4 \\
LSBVCC299 &  12  24  30.9 &  11  48  42.1 &  18.5 &  23.8 &  4.6 \\
LSBVCC300 &  12  24  56.1 &  11  49  56.3 &  17.2 &  22.6 &  4.7 \\
LSBVCC301 &  12  24  47.9 &  11  49   6.6 &  19.4 &  24.1 &  3.5 \\
LSBVCC302 &  12  23  57.8 &  11  53  31.9 &  18.0 &  23.3 &  4.5 \\
LSBVCC303 &  12  23  38.9 &  12  11   2.0 &  18.6 &  23.0 &  3.1 \\
LSBVCC304 &  12  22  53.8 &  12  23  20.0 &  19.5 &  24.3 &  3.6 \\
LSBVCC305 &  12  32   1.4 &   8  40  10.9 &  17.8 &  22.9 &  4.3 \\
LSBVCC306 &  12  30  46.0 &   9   0  45.4 &  18.8 &  23.6 &  3.8 \\
LSBVCC307 &  12  33  51.0 &   9   4  45.5 &  21.3 &  25.7 &  3.0 \\
LSBVCC308 &  12  31  16.1 &   9  21  33.1 &  19.2 &  23.9 &  3.6 \\
LSBVCC309 &  12  30  16.2 &   9  27  43.6 &  20.4 &  24.7 &  3.0 \\
LSBVCC310 &  12  31  24.7 &   9  28  28.2 &  18.8 &  23.4 &  3.3 \\
LSBVCC311 &  12  45   9.7 &  10  37  49.1 &  17.3 &  23.8 &  7.7 \\
LSBVCC312 &  12  43  35.4 &  10  45  20.5 &  18.1 &  22.6 &  3.2 \\
LSBVCC313 &  12  45  25.1 &  10  55  26.0 &  18.9 &  23.8 &  3.9 \\
LSBVCC314 &  12  44  54.6 &  10  57  41.8 &  19.2 &  24.1 &  3.7 \\
LSBVCC315 &  12  42   2.8 &  10  57  13.3 &  18.1 &  24.2 &  6.7 \\
LSBVCC316 &  12  44  54.2 &  11   1   8.0 &  19.0 &  24.0 &  3.9 \\
LSBVCC317 &  12  44  16.8 &  11  15  32.4 &  19.4 &  24.9 &  4.9 \\
LSBVCC318 &  12  44  31.5 &  11  23  54.2 &  20.1 &  24.9 &  3.6 \\
LSBVCC319 &  12  43  48.7 &  11  25  31.1 &  19.9 &  24.6 &  3.6 \\
LSBVCC320 &  12  43  18.0 &  11  28  27.5 &  18.6 &  24.2 &  5.3 \\
LSBVCC321 &  12  45  35.4 &  11  33   7.6 &  18.4 &  23.1 &  3.5 \\
LSBVCC322 &  12  43   8.1 &  11  37  16.7 &  19.4 &  24.3 &  3.8 \\
LSBVCC323 &  12  43   2.2 &  11  41  52.1 &  16.8 &  22.9 &  6.7 \\
LSBVCC324 &  12  42  22.9 &  11  41   7.1 &  19.6 &  24.6 &  4.1 \\
LSBVCC325 &  12  44  33.7 &  11  47  43.1 &  17.7 &  23.2 &  4.9 \\
LSBVCC326 &  12  44  44.9 &  11  48   4.0 &  19.0 &  24.5 &  4.9 \\
LSBVCC327 &  12  43  57.8 &  11  52  50.5 &  18.1 &  23.8 &  5.6 \\
LSBVCC328 &  12  44  53.0 &  12  10  58.8 &  18.4 &  23.0 &  3.4 \\
LSBVCC329 &  12  45  45.1 &  12  14   8.2 &  18.0 &  22.6 &  3.3 \\
LSBVCC330 &  12  45   4.0 &  12  21   5.4 &  17.0 &  22.9 &  5.9 \\
LSBVCC331 &  12  42   3.2 &  12  28  49.8 &  18.0 &  23.6 &  5.2 \\
LSBVCC332 &  12  30  23.4 &  15  41  58.6 &  17.2 &  23.0 &  5.8 \\
LSBVCC333 &  12  33  35.2 &  16  17  57.1 &  18.4 &  24.3 &  6.2 \\
LSBVCC334 &  12  33  30.2 &  16  17  41.3 &  18.2 &  22.6 &  3.0 \\
LSBVCC335 &  12  31  35.9 &  16  43  29.6 &  18.6 &  23.2 &  3.2 \\
LSBVCC336 &  12  31  58.4 &  17   1  23.9 &  18.1 &  23.7 &  5.2 \\
LSBVCC337 &  12  30  31.2 &   6  23  19.3 &  19.8 &  24.8 &  3.9 \\
LSBVCC338 &  12  31  13.7 &   6  40  16.7 &  18.5 &  24.0 &  5.0 \\
LSBVCC339 &  12  31  53.9 &   5  10  23.2 &  18.4 &  23.5 &  4.1 \\
LSBVCC340 &  12  31  50.4 &   5  28  58.1 &  19.1 &  24.9 &  5.9 \\
LSBVCC341 &  12  31   0.6 &   5  33  19.1 &  18.1 &  23.2 &  4.2 \\
LSBVCC342 &  12  33  47.4 &   5  33  29.9 &  18.9 &  24.5 &  5.1 \\
LSBVCC343 &  12  33  55.7 &   5  43   8.8 &  18.1 &  22.9 &  3.7 \\
LSBVCC344 &  12  34   0.8 &   5  50  28.0 &  20.0 &  24.5 &  3.3 \\
LSBVCC345 &  12  13  35.7 &  13   2   4.6 &  17.6 &  23.4 &  5.6 \\
LSBVCC346 &  12  12  29.2 &  13   3  32.4 &  20.9 &  25.4 &  3.1 \\
LSBVCC347 &  12  12  29.5 &  13   9  59.8 &  20.0 &  25.1 &  4.2 \\
LSBVCC348 &  12  14   9.8 &  13  14   0.6 &  19.4 &  23.7 &  3.0 \\
LSBVCC349 &  12  12  26.8 &  13  16  46.2 &  18.3 &  23.4 &  4.1 \\
LSBVCC350 &  12  11  21.4 &  13  17   6.0 &  20.7 &  25.3 &  3.4 \\
LSBVCC351 &  12  11  28.3 &  13  35   0.6 &  18.5 &  23.0 &  3.2 \\
LSBVCC352 &  12  11  53.7 &  14  46  34.3 &  17.8 &  22.8 &  3.9 \\
LSBVCC353 &  12  14   1.1 &  14  48  11.2 &  20.4 &  25.5 &  4.3 \\
LSBVCC354 &  12  14  22.2 &  14  58  43.3 &  17.8 &  22.9 &  4.2 \\
LSBVCC355 &  12   8   5.4 &  12  45  50.4 &  17.7 &  23.7 &  6.2 \\
LSBVCC356 &  12  10  25.0 &  13  21  58.7 &  18.5 &  23.3 &  3.6 \\
LSBVCC357 &  12  21  23.4 &  13  35   0.2 &  20.3 &  24.7 &  3.0 \\
LSBVCC358 &  12  22  22.6 &  14  25  49.1 &  20.3 &  25.2 &  3.7 \\
LSBVCC359 &  12  19  47.7 &  14  42  23.4 &  18.2 &  23.3 &  4.1 \\
LSBVCC360 &  12  20  26.9 &  14  47   7.4 &  17.3 &  22.9 &  5.2 \\
LSBVCC361 &  12  22  20.9 &  15   9  36.0 &  18.8 &  23.5 &  3.4 \\
LSBVCC362 &  12  22   9.6 &  15  39  10.4 &  19.1 &  25.0 &  6.0 \\
LSBVCC363 &  12  19  50.4 &  15  40  20.3 &  17.9 &  23.4 &  5.0 \\
LSBVCC364 &  12  20  24.8 &  16   4  21.4 &  19.3 &  23.7 &  3.1 \\
LSBVCC365 &  12  19  50.6 &  16  16   7.3 &  18.8 &  23.2 &  3.0 \\
LSBVCC366 &  12  20  51.4 &  16  21  49.0 &  19.1 &  24.9 &  5.8 \\
LSBVCC367 &  12  20  14.8 &  16  34   6.6 &  20.4 &  25.0 &  3.3 \\
LSBVCC368 &  12  20  29.6 &  16  48  56.5 &  20.3 &  25.1 &  3.7 \\
LSBVCC369 &  12  22  12.3 &  16  58  29.3 &  18.7 &  23.9 &  4.3 \\
LSBVCC370 &  12  21  36.5 &   5  49  23.9 &  18.2 &  23.1 &  3.8 \\
LSBVCC371 &  12  21  17.1 &   5  59  44.9 &  19.5 &  24.5 &  3.9 \\
LSBVCC372 &  12  20  17.8 &   6   0   5.8 &  20.0 &  24.7 &  3.5 \\
LSBVCC373 &  12  21  38.4 &   6  16  59.5 &  19.3 &  25.1 &  5.9 \\
LSBVCC374 &  12  20  52.9 &   6  19  28.9 &  18.0 &  22.6 &  3.4 \\
LSBVCC375 &  12  19  58.6 &   6  20   3.8 &  18.6 &  23.6 &  4.0 \\
LSBVCC376 &  12  21  22.3 &   6  23  38.8 &  18.5 &  23.4 &  3.9 \\
LSBVCC377 &  12  16  23.2 &   7  47  58.9 &  17.2 &  23.3 &  6.7 \\
LSBVCC378 &  12  16  50.1 &   7  51  45.0 &  18.2 &  22.8 &  3.3 \\
LSBVCC379 &  12  16  33.7 &   6  50   5.6 &  17.6 &  22.6 &  4.0 \\
LSBVCC380 &  12  16  53.2 &   6  54  41.0 &  17.4 &  22.8 &  4.7 \\
LSBVCC381 &  12  16  59.2 &   7   1  12.4 &  19.7 &  24.2 &  3.2 \\
LSBVCC382 &  12  17   1.0 &   7   6  41.4 &  19.8 &  24.4 &  3.4 \\
LSBVCC383 &  12  17  30.8 &   7  19  20.3 &  20.6 &  25.0 &  3.1 \\
LSBVCC384 &  12  17   7.2 &   7  44   0.2 &  17.4 &  22.5 &  4.1 \\
LSBVCC385 &  12  31   4.6 &  17  23  16.4 &  18.2 &  23.8 &  5.1 \\
LSBVCC386 &  12  34  14.9 &  17  43   0.8 &  20.6 &  25.0 &  3.0 \\
LSBVCC387 &  12  31  23.2 &  17  55   7.7 &  18.5 &  23.2 &  3.6 \\
LSBVCC388 &  12  35  31.8 &   7   1  35.0 &  18.0 &  23.0 &  3.9 \\
LSBVCC389 &  12  37  41.1 &   7  40  37.2 &  17.8 &  23.7 &  6.1 \\
LSBVCC390 &  12  40  21.3 &  12  43   8.8 &  19.2 &  24.4 &  4.4 \\
LSBVCC391 &  12  50  52.5 &  13  41  41.6 &  18.4 &  22.8 &  3.1 \\
LSBVCC392 &  12  50  18.5 &  14   5  52.4 &  18.5 &  24.0 &  4.9 \\
LSBVCC393 &  12  29   2.1 &  12  26   7.1 &  19.0 &  23.8 &  3.6 \\
LSBVCC394 &  12  29  28.6 &  12  29  46.7 &  17.5 &  22.7 &  4.4 \\
LSBVCC395 &  12  27  49.5 &  12  29  58.6 &  19.6 &  24.7 &  4.0 \\
LSBVCC396 &  12  26  20.1 &  12  30  37.4 &  18.1 &  22.6 &  3.2 \\
LSBVCC397 &  12  30  15.2 &  12  30  56.5 &  19.0 &  23.8 &  3.6 \\
LSBVCC398 &  12  28  42.6 &  12  33   0.0 &  16.9 &  23.0 &  6.7 \\
LSBVCC399 &  12  27   2.7 &  12  34  49.8 &  19.6 &  25.1 &  5.1 \\
LSBVCC400 &  12  29  53.7 &  12  37  15.2 &  19.3 &  24.7 &  4.7 \\
LSBVCC401 &  12  27  15.7 &  12  39  36.4 &  19.1 &  25.4 &  7.1 \\
LSBVCC402 &  12  28  47.1 &  12  38  30.8 &  19.9 &  24.4 &  3.1 \\
LSBVCC403 &  12  26  52.1 &  12  39   8.6 &  20.2 &  24.7 &  3.2 \\
LSBVCC404 &  12  28  55.5 &  12  42  24.8 &  18.6 &  23.3 &  3.4 \\
LSBVCC405 &  12  26  27.9 &  12  45  51.5 &  20.7 &  25.4 &  3.4 \\
LSBVCC406 &  12  26  36.5 &  12  48  11.9 &  18.1 &  23.6 &  5.1 \\
LSBVCC407 &  12  27  34.4 &  12  48  14.0 &  19.5 &  24.8 &  4.7 \\
LSBVCC408 &  12  27  39.0 &  12  52  47.6 &  17.6 &  23.8 &  6.7 \\
LSBVCC409 &  12  27  14.3 &  12  54   9.0 &  20.1 &  25.0 &  3.7 \\
LSBVCC410 &  12  29  46.3 &  12  55  27.8 &  20.7 &  25.0 &  3.0 \\
LSBVCC411 &  12  28   6.8 &  12  58  43.0 &  18.8 &  23.7 &  3.8 \\
LSBVCC412 &  12  26  38.3 &  13   4  46.9 &  19.0 &  24.5 &  5.2 \\
LSBVCC413 &  12  26  46.7 &  13  16   2.6 &  19.1 &  24.3 &  4.5 \\
LSBVCC414 &  12  28  19.9 &  13  21  34.9 &  20.5 &  25.0 &  3.1 \\
LSBVCC415 &  12  29  20.8 &  13  22  12.0 &  20.9 &  25.3 &  3.0 \\
LSBVCC416 &  12  27  15.6 &  13  24  46.1 &  19.9 &  24.8 &  3.9 \\
LSBVCC417 &  12  27  35.5 &  15  25  42.2 &  20.1 &  24.8 &  3.5 \\
LSBVCC418 &  12  28  57.8 &  15  34  58.8 &  19.3 &  24.1 &  3.7 \\
LSBVCC419 &  12  28  20.2 &  15  41  59.6 &  18.5 &  23.8 &  4.4 \\
LSBVCC420 &  12  29  30.8 &  15  44   3.5 &  17.8 &  22.6 &  3.7 \\
LSBVCC421 &  12  29  17.4 &  15  44  56.4 &  20.1 &  25.1 &  4.0 \\
LSBVCC422 &  12  28  31.5 &  16   3  51.5 &  17.9 &  22.8 &  3.8 \\
LSBVCC423 &  12  26  52.3 &  16  15   1.4 &  19.9 &  25.1 &  4.3 \\
LSBVCC424 &  12  26  30.1 &  16  15  53.6 &  20.0 &  24.8 &  3.7 \\
LSBVCC425 &  12  26  46.9 &  16  16  37.2 &  20.8 &  25.3 &  3.1 \\
LSBVCC426 &  12  26  34.2 &  16  27  51.1 &  19.7 &  25.3 &  5.2 \\
LSBVCC427 &  12  29   0.5 &  16  42   4.7 &  17.9 &  23.5 &  5.2 \\
LSBVCC428 &  12  30   8.9 &   9  42  57.2 &  19.4 &  25.0 &  5.2 \\
LSBVCC429 &  12  29  59.1 &   9  47  48.1 &  19.2 &  24.6 &  5.0 \\
LSBVCC430 &  12  30  14.1 &  10  10  51.6 &  17.2 &  23.7 &  7.8 \\
LSBVCC431 &  12  29  25.9 &  10  14  57.1 &  17.8 &  23.2 &  4.7 \\
LSBVCC432 &  12  30   8.1 &  10  15  52.6 &  19.9 &  24.4 &  3.1 \\
LSBVCC433 &  12  28  12.0 &  10  21  52.6 &  16.7 &  23.1 &  7.7 \\
LSBVCC434 &  12  29  18.2 &  10  29  48.5 &  18.0 &  22.5 &  3.1 \\
LSBVCC435 &  12  28  31.4 &  10  31   8.4 &  17.3 &  22.9 &  5.3 \\
LSBVCC436 &  12  28  12.9 &  10  31  33.2 &  17.0 &  23.0 &  6.2 \\
LSBVCC437 &  12  26  41.8 &   4  54  23.8 &  19.5 &  24.6 &  4.1 \\
LSBVCC438 &  12  22  10.5 &  11  38  30.1 &  18.4 &  23.1 &  3.5 \\
LSBVCC439 &  12  21  48.9 &  11  51   8.6 &  19.0 &  23.5 &  3.2 \\
LSBVCC440 &  12  18  53.8 &  12  11  39.5 &  18.6 &  23.9 &  4.6 \\
LSBVCC441 &  12  19  30.7 &  12  12  31.7 &  18.4 &  23.3 &  3.8 \\
LSBVCC442 &  12  20  55.0 &  12  19  45.1 &  19.0 &  24.0 &  4.0 \\
LSBVCC443 &  12  21  15.3 &  12  24  24.5 &  18.8 &  24.0 &  4.4 \\
        
\caption{The LSBVCC galaxy sample - (1) name, (2) (3) position, (4) $g$ band apparent magnitude, (5) $g$ band central surface brightness (6) $g$ band exponential scale length.}
\end{longtable}
\end{center}

\end{document}